\documentclass{article}

\usepackage{PRIMEarxiv}

\usepackage[utf8]{inputenc} 
\usepackage[T1]{fontenc}    
\usepackage{hyperref}       
\usepackage{url}            
\usepackage{booktabs}       
\usepackage{amsfonts}       
\usepackage{nicefrac}       
\usepackage{microtype}      
\usepackage{lipsum}
\usepackage{fancyhdr}       
\usepackage{graphicx}       
\usepackage{amssymb}
\usepackage{algorithm}
\graphicspath{{media/}}     
\usepackage{amsmath}
\title{Hybrid Quantum Algorithms integrating QAOA, Penalty Dephasing and Zeno Effect for Solving Binary Optimization Problems with Multiple Constraints
\thanks{\textit{\underline{FinQ Tech Inc.}}: 
\textbf{Ke Wan, Yiwen Liu:Hybrid Quantum Algorithms integrating QAOA, Penalty Dephasing and Zeno Effect for Solving Binary Optimization Problems with Multiple Constraints}} 
}

\author{
  Ke Wan \\
  Princeton Univ \\
  \texttt{kwan@alumni.princeton.edu} \\
   \And
  Yiwen Liu  \\
  Ohio State Univ \\
  \texttt{liuyiwen153@gmail.com} \\
  \And
}

\begin{document}
\maketitle

\begin{abstract}

When tackling binary optimization problems using quantum algorithms, the conventional Ising representation and Quantum Approximate Optimization Algorithm (QAOA) encounter difficulties in efficiently handling errors for large-scale problems involving multiple constraints. To address these challenges, this paper presents a hybrid framework that combines the use of standard Ising Hamiltonians to solve a subset of the constraints, while employing non-Ising formulations to represent and address the remaining constraints. The resolution of these non-Ising constraints is achieved through either penalty dephasing or the quantum Zeno effect. This innovative approach leads to a collection of quantum circuits with adaptable structures, depending on the chosen representation for each constraint. Furthermore, this paper introduces a novel technique that utilizes the quantum Zeno effect by frequently measuring the constraint flag, enabling the resolution of any optimization constraint. Theoretical properties of these algorithms are discussed, and their performance in addressing practical aircraft loading problems is highly promising, showcasing significant potential for a wide range of industrial applications.
\end{abstract}

\keywords{Quantum Computing \and QAOA \and Quantume ZENO Effect \and AOC \and Penalty Dephasing \and Binary Optimization}

\section{Introduction}

Quantum computing offers a novel approach to solving real-world optimization problems by utilizing the state potentials of a quantum system to represent the objective function and potential solutions as superposition of quantum states. This differs significantly from classical approaches that rely on neighborhood searching and descent methods. One of the major advantages of quantum computing is that it can explore all potential solutions via quantum superposition in each step, whereas classical algorithms can only explore one or a limited number of solutions each step. As a result, quantum optimization algorithms greatly enhance the efficiency of exploring problem structures. Moreover, solving optimization problems using a quantum computing system leverages various operations in quantum mechanics, which lead to innovative algorithms.

In this paper, we review several alternative methods for designing quantum circuits that can be utilized to solve optimization problems with constraints. We integrate these methods into a hybrid framework to solve binary optimization problems with multiple constraints, including the standard Quantum Approximate Optimization Algorithm (QAOA), Quantum Dephasing, and Quantum ZENO effect. The basics of each method are reviewed in the next few sections to provide a foundation for future research.

\subsection{Review of Classical Methods for Constrained Optimization} 

To provide more context from the optimization literature, the classical way of resolving a constraint optimization problem is summarized below:

Boyd's book on convex optimization  \cite{boyd2004convex} provides a comprehensive framework for solving convex optimization problems, which are a subset of optimization problems where the objective function and inequality constraint functions are all convex, and equality constraint functions are affine. To solve constrained optimization problems, Boyd suggests augmenting the objective function with a weighted sum of the constraints using Lagrange multipliers.

In recent years, the alternating direction method of multipliers (ADMM) has emerged as a popular algorithm for solving convex optimization problems \cite{boyd2011distributed}\cite{parikh2014proximal}. ADMM breaks the problem down into smaller subproblems that are easier to handle, and updates the variables separately while maintaining consensus between them. This approach can be particularly effective for problems with complex constraints.

Boyd(2004) \cite{boyd2004convex} set up the framework of convex optimization in his book, While an unconstrained optimization is solved more easily using numerical optimization problem, the objective function of a constrained optimization problems is augmented with a weighted sum of the constraint using Lagrange multipliers.  A convex optimization is defined as an optimization problem whose objective function, inequality constraint functions are all convex, equality constraint functions are affine. More recently, alternating direction method of multipliers (ADMM) is proposed as an algorithm that solves convex optimization problems by breaking them into smaller pieces, each of which are then easier to handle \cite{boyd2011distributed}\cite{parikh2014proximal}. 

For a convex optimization problem, numerical optimization methods are generally reliable in converging to the optimal solution. However, non-convex optimization problems can pose greater challenges for numerical optimization. The algorithm's solution path can be categorized as an interior point method if it remains within the feasible region throughout the optimization process, or an exterior point method if it ventures outside the feasible region.

As summarized in Freund(2004) \cite{freund2004penalty}
Penalty methods and barriers method are used to solve constraint optimization.
\begin{itemize}
 \item In a penalty method, the original feasible region is expanded to $R^n$ , but a large cost or “penalty” is added to the objective function for points that lie outside of the original feasible region. Coralia (2018)\cite{Coralianum} discussed how to derive a sequence of perturbed KKT conditions to solve the constraint optimization using numerical searching.
 \item In a barrier method, we impose a very large cost on feasible points that lie ever closer to the boundary of the feasible region, thereby creating a “barrier” to exiting the feasible region. Coralia (2018)\cite{Coralianum} introduced log penalty and uses central paths and trust regions to solve constraint optimization.
\end{itemize}

Binary optimization is a specific sub-category of optimization problems where the decision variables are 0 or 1, which represents a wide class of problem where decisions are made between two states. Usually it can be solved as a integer programming using branch and cut.  
Yuan(2016) \cite{yuan2016binary} proposed a new class of continuous optimization techniques which is based on Mathematical Programming with Equilibrium Constraints (MPECs), in which they first reformulate the binary program as an equivalent augmented biconvex optimization problem with a bilinear equality constraint, then they propose two penalization/regularization methods (exact penalty and alternating direction) to solve it.

Guan(2017) \cite{yuan2017exact}  presented a reformulation of the binary program as an augmented biconvex optimization problem with a bilinear equality constraint, followed by proposing an exact penalty method to solve it. The resulting algorithm aims to find an optimal solution to the original problem by solving a sequence of linear programming convex relaxation sub-problems.

Binary optimization problems have a natural connection to the Ising Hamiltonian, which will be introduced in the next section. This makes them a class of problems that quantum circuits can solve efficiently. In this paper, we consider a binary optimization problem with convex objective and multiple constraints; several alternative quantum algorithms are used to solve it while a classical simulated annealing method are used as benchmark.

\subsection{Review of Basic Concepts in Quantum Computing}

Basic terminologies of quantum computing are reviewed in this section to facilitate discussions in this paper \cite{messiah2014quantum} \cite{QuantumCir_IBM} \cite{QuantumCir_Wiki} \cite{QuantumCir_britannica}: 
\begin{itemize}
\item Quantum bit (qubit): A qubit is a two-state (or two-level) quantum-mechanical system, one of the simplest quantum systems displaying the peculiarity of quantum mechanics. 
\item Classical bit: A bit in a classical system, which would have to be in one state or the other. 
\item Quantum gates:a quantum logic gate is a basic quantum circuit operating on a small number of qubits. They are the building blocks of quantum circuits, like classical logic gates are for conventional digital circuits.
\item Quantum circuit:A quantum circuit is a computational routine consisting of coherent quantum operations on quantum data, such as qubits, and concurrent real-time classical computation. 
\item Superposition: It is a fundamental principle of quantum mechanics, which describes any two (or more) quantum states can be added together ("superposed") and the result will be another valid quantum state. 
\item Hamiltonian: The Hamiltonian of a system specifies its total energy—i.e., the sum of its kinetic energy (that of motion) and its potential energy (that of position)—in terms of the Lagrangian function derived in earlier studies of dynamics and of the position and momentum of each of the particles.
\end{itemize}

Quantum circuits can solve optimization problems through a class of algorithms called quantum optimization algorithms. There are two main stream of algorithms 
\begin{itemize}
\item Quantum annealing: In quantum annealing, the optimization problem is mapped onto a problem of finding the ground state of a corresponding Ising spin glass Hamiltonian. The Hamiltonian encodes the objective function and constraints of the optimization problem. The optimization problem can then be solved by preparing the quantum anneal in its ground state, which corresponds to the optimal solution of the optimization problem.
\item Quantum approximate optimization algorithms (QAOA), which are implemented using general-purpose quantum computers. QAOA is a variational algorithm that uses quantum circuits to prepare a trial solution, which is then optimized using classical optimization techniques. QAOA has been shown to provide a quantum speedup for certain optimization problems, although its performance is not yet competitive with classical optimization methods for all problems. 
\end{itemize}

\subsection{Review of the QAOA Algorithm}

This section provides a detailed review of the Quantum Approximate Optimization Algorithm (QAOA). QAOA is a hybrid quantum-classical algorithm widely used to find approximate solutions to combinatorial optimization problems. The algorithm involves using a series of quantum gates to prepare a quantum state, which is then measured and used to calculate the objective function value of an optimization problem. This process is repeated multiple times to search for the optimal solution.

QAOA is based on several concepts, including Variational Quantum Eigensolver (VQE), Adiabatic Quantum Computation (AQC), and Ising Model.


\begin{itemize}
\item VQE \cite{peruzzo2014variational} and \cite{tilly2022variational} are algorithms designed to find a parameterization that minimizes the expectation value of a Hamiltonian. The goal is to approximate the eigenvector of the Hermitian operator corresponding to the lowest eigenvalue. This expectation value provides an upper bound on the ground state energy and, ideally, should be indistinguishable from it to the desired level of precision. QAOA can be seen as an extension of VQE, where the Hamiltonian to be minimized is replaced by the objective function of the optimization problem.

\item Adiabatic Optimization (AOC) is a quantum computational technique that involves slowly changing the Hamiltonian of a quantum system from an initial Hamiltonian to a final Hamiltonian that encodes the optimization problem \cite{aharonov2008adiabatic}\cite{albash2018adiabatic}. Initially, the system is prepared in the ground state of the initial Hamiltonian. By slowly changing the Hamiltonian to the final Hamiltonian, the system remains in the ground state throughout the process, thereby finding the optimal solution. AQC has demonstrated effectiveness in solving various optimization problems, including Ising spin glasses and satisfiability problems.

\item The Ising model is a fundamental model for interacting systems, in which a large number of spins are individually in microscopic states of +1 or -1, with the value of a spin at a site determined by its (usually short-range) interaction with the other spins in the model. This model can be extended to a system of n qubits, where the spin-down and spin-up states can be represented by the qubit states $|0>$ and $|1>$, corresponding to energy labels of -1 and +1. These energy levels of a qubit can act as the eigenvalues of a Hamiltonian corresponding to the eigenstates $|0>$ and $|1>$. Binary optimization problems are closely linked to the Ising representation through a linear mapping of variables from $[-1,1]$ to $[0,1]$. This mapping allows binary optimization problems to be transformed into Ising Hamiltonian problems. Quantum circuits can be created based on the Ising Hamiltonian, but typically require a large number of gates.

\end{itemize}

QAOA is used to solve optimization problems by defining an objective function in terms of a Hamiltonian Hc, whose expectation value <Hc> is the quantity to be optimized. The minimum eigenvalue of the matrix Hc corresponds to the minimum expectation value $<Hc>$, which is the optimal solution we seek. Similarly, the maximum expectation value $<Hc>$ corresponds to the maximum eigenvalue of Hc.

To determine the required eigenstate for a given Hamiltonian expectation value, QAOA leverages adiabatic computing. This involves evolving the quantum state of the system from an initial Hamiltonian Ho to the final Hamiltonian of interest. We can achieve this by evolving the quantum system in p steps using unitary transformations. The quality of the approximation improves as p is increased. The quantum circuit that implements QAOA consists of unitary gates whose locality is at most the locality of the objective function being optimized. The depth of the circuit grows linearly with $p$ times (at worst) the number of constraints.



Lucas (2014) \cite{lucas2014ising} provided Ising formulations for many NP-complete and NP-hard problems, extending mappings to the Ising model from partitioning, covering, and satisfiability.
Crooks (2018) \cite{crooks2018performance}  studied the performance of QAOA on the maxcut problem and found that it requires $O(N^2P)$ gates and has a run time of O(NP), assuming O(N) gates are applied in parallel. Zhou (2020) provided an in-depth study of QAOA's performance on the maxcut problem, introducing an efficient parameter estimation procedure in $O(poly(p))$ time where p is the layer of the QAOA network.
Franze (2022) \cite{fuchs2022constrained} presented a framework for constructing mixing operators that restrict the evolution to a subspace of the full Hilbert space given by these constraints. They also described algorithms for efficient decomposition into basis gates. Parer concluded that while designing mixers with the presented framework is more or less straightforward, designing efficient mixers turns out to be a difficult task. 

Please note that if using classical QAOA algorithm to solve constrained optimization, one has to represent the problem objective and the constraints together in Lagrange form, which is then converted into an integrated circuits.

\subsection{Review of Quantum Zeno Effect}

The quantum Zeno effect, also known as the Turing paradox, is a characteristic of quantum-mechanical systems in which continuous measurement or observation causes a quantum system to remain in its initial state and not evolve. This effect can slow down a particle's time evolution by frequently measuring it with respect to a chosen measurement setting, thereby hindering it from transitioning to states different from its initial state. The effect is named after the ancient Greek philosopher Zeno of Elea, who proposed a paradox involving an arrow in flight that appeared to be motionless due to a series of infinitely short intervals of time. The Zeno effect is a consequence of general features of the Schrödinger equation and is unique to quantum systems, which may produce unique solutions when applied to various problems. More details on this phenomenon can be found in the appendix.
The quantum Zeno effect has been extensively studied in the literature. Facci (2009) \cite{facchi2008quantum} summarizes three ways of achieving the Zeno effect: projective measurements, unitary kicks, and a strong continuous coupling. They also introduce the concept of Quantum Zeno subspaces through blockwise decomposition. Burgarth (2020) \cite{burgarth2020quantum} considers the evolution of a finite-dimensional quantum system under frequent kicks, where each kick is a generic quantum operation. They develop a generalization of the Baker-Campbell-Hausdorff formula, which reveals an adiabatic evolution. Herman (2022) \cite{herman2022portfolio} proposes a technique that uses quantum Zeno dynamics to solve optimization problems with multiple arbitrary constraints, including inequalities. They demonstrate that the dynamics of the quantum optimization can be efficiently restricted to the in-constraint subspace via repeated projective measurements. To address the limitations of adiabatic quantum computation, Yu (2022) \cite{yu2021quantum}  adopts a Zeno method, where a series of eigenstate projections are used in the quantum simulated annealing. The path-dependent Hamiltonian is augmented by a sum of Pauli X terms, whose contribution vanishes at the beginning and the end of the path. More details on these works can be found in the Appendix.


In this paper, we propose a formal application of the quantum Zeno effect to solve optimization problems with multiple constraints. This approach greatly expands the horizon for its applications. The concept of the quantum Zeno effect allows the total Hilbert space to be divided into Zeno subspaces, and different components of the density matrix can independently evolve within each sector. By constantly measuring the system, the initial state of one of the subspaces would have a survival probability of unity. We can expand this idea into the Quantum Approximate Optimization Algorithm (QAOA) by restricting the system evolution only within the subspace where the feasible solution resides, rather than introducing all the constraints into the target function. To achieve this approach, we need to initialize the system evolution from only feasible states and introduce the Zeno effect by performing frequent projective measurements within the mixer. The purpose is to minimize the probability that the mixing operator would transition from a feasible state to an infeasible state, ideally approaching zero.

\subsection{Review of Penalty Dephasing}

The third widely studied pattern of quantum circuit is the quantum penalty dephasing algorithm, which has proven to be innovative in solving non-Ising target functions in quantum computing. This algorithm is rooted in Adiabatic Quantum Computation (AQC) and adiabatic approximation. AQC involves slowly changing the Hamiltonian of a quantum system from an initial Hamiltonian to a final Hamiltonian that encodes the optimization problem \cite{aharonov2008adiabatic}\cite{albash2018adiabatic}. The system is initially prepared in the ground state of the initial Hamiltonian and remains in the ground state throughout the process, thereby finding the optimal solution. AQC has been effective in solving various optimization problems, including Ising spin glasses and satisfiability problems.

Any optimization problem can be reformulated for use in Adiabatic Quantum Computation (AQC), where a Hamiltonian is constructed whose ground state encodes the solution to the optimization problem. The quantum system is then prepared in the ground state of a simple initial Hamiltonian, and the system evolves adiabatically under a time-dependent Hamiltonian to reach the final Hamiltonian, which encodes the optimal solution. One way to implement the time-dependent Hamiltonian is to model it as a linear combination of the initial and final Hamiltonians, with time-dependent coefficients $\beta_t$ and $\gamma_t$, respectively. These coefficients can be varied through time such that $\beta_t$ decreases and $\gamma_t$ increases. To convert the AQC into a quantum circuit, a discrete time scheme is introduced, breaking down the time evolution into multiple steps. At each step, the time evolution of a state under a time-independent Hamiltonian can be expressed using the unitary time evolution operator. The overall evolution is then expressed as a product of such unitary evolution operators.


For relevant research, Farhi (2014) \cite{farhi2014quantum} proposed a quantum algorithm for generating approximate solutions to combinatorial optimization problems. The algorithm's accuracy improves with an increasing integer p. The quantum circuit that implements the algorithm consists of unitary gates whose locality is no more than that of the objective function for which the optimal value is sought. Grand'rive (2019) \cite{de2019knapsack} introduced a relaxed algorithm for solving constrained optimization problems with non-standard Ising target functions. This approach adds penalties to solutions that do not meet the constraints, which is implemented through separate "cost addition" and "constraint testing and penalty dephasing" blocks rather than within the classical Ising Hamiltonian. Lukas (2021) \cite{lucas2021ibmsolution} developed a robust implementation based on this literature.

To enable such circuits, quantum networks that perform elementary arithmetic operations, including modular exponentiation, have been introduced by Vedral (1995) \cite{vedral1996quantum}. In addition, various quantum adders have been introduced, such as the quantum ripple-carry addition circuit by Cuccaro (2004) \cite{cuccaro2004new}, the quantum Fourier transform adder by Draper (2000) \cite{draper2000addition}, and the adder by Ruiz (2014) \cite{ruiz2017quantum}. The modular exponentiation dominates the overall time and memory complexity in Shor's quantum factorization algorithm.

\section{Combine Zeno, Dephasing and QAOA for a Optimization Problem with Multiple Constraints}

In this section, the framework which assembled Quantum ZENO, Penalty Dephasing and standard QAOA for an optimization problem with multiple constraints are proposed.

\subsection{The Hybrid Framework}
Given an optimization problem with multiple constraints, both Ising and non-Ising formulation can be employed to represent part of the problem. For the part which is represented in Ising Hamiltonian, one applies standard QAOA algorithm to solve it; for the non-Ising part, one can further apply either Quantum Zeno Effect or Penalty Dephasing. Such hybrid method is an innovative way of solving optimization through quantum computing. Detailed descriptions are provided below:

\begin{eqnarray}
minimize f(x)\\
 s.t.  
 g_i(x)< 0 
 \end{eqnarray}

\subsubsection{The framework as an optimization problem}
Suppose there are $N=m+n+k$ constraints in total, and we propose the $N$ constraints can be represented in three forms:
\begin{itemize}
\item $m$ constraints are represented in Standard Ising Hamiltonian
This Hamiltonian is linked to the Lagrange function
\begin{eqnarray}f(x)+\lambda_i (g_i(x)-\delta_i)^2 \end{eqnarray}
for $i \in G_m$ 
\item $n$ constraints are represented in non-Ising Hamiltonian - Zeno
\item $k$ constraints are represented through non-Ising Hamiltonian - Dephasing
\end{itemize}

When converting to quantum circuit, the following rule is applied
\begin{itemize}
\item All standard Ising are represented in a standard Ising circuit as base circuit.
\item To consider one non-Ising formulated constraint, one additional layer is added; each of such layers can follow either Quantum Zeno Effect or Penalty Dephasing. Such layered structure aligns with the general flow of quantum circuit and the outcome of internal layers are treated as 'phase return' to outside layers. \item For constraints that are tackled through zeno circuit, we start with initial feasible state so that zeno-constraint subspace is feasible. Mathematically, the solution is restricted in the feasible subspace $V$, corresponding to $C_i$, which are represented by ZENO. Please note this can be a bigger space than the full feasible space of the original problem.  
\end{itemize}  

Overall, different solving logic is assembled in a layered structure which is displayed in figure below. The two subsections below will devote to describe the design on the two layers for Quantum ZENO effect and Penalty Dephasing respectively. 

\subsubsection{The framework as a quantum computing problem}
From quantum perspective, a universal theoretical framework to integrate the three quantum circuits with different characteristics. We start by defining expanded unitary as \begin{eqnarray}U(z)|z>=U_o(z)*U_z(z)*U_d(z)|z>\end{eqnarray}. Each block of which is discussed in further details below. Please note $i$ denotes the symbol for imaginary number, while $j$ are used as the indexer. 
\begin{itemize}
\item Denote the set of all constraints that are solved in QAOA as $C_q$. The part corresponding to the original objective function and these constraints are expressed with Hamiltonian below
\begin{eqnarray}
H_o&=&f(x)+\sum_{j \in C_1}\lambda_j (g_j(x)-\delta_j)^2
\end{eqnarray} 
with the corresponding unitary 
\begin{eqnarray}
U_o(z)|z>\, &=& e^{i\theta H_o(z)}|z>\, \\
&=& e^{-i\theta (f(z)+\sum_{j \in C_q}\lambda_j (g_j(z)-\delta_j)^2)}|z>
\end{eqnarray}  

In more details, the following transformation is conducted:
\begin{itemize}
\item We introduced slack variables $\delta_i$ to translate them into equality constraint and use L2 norm to avoid introducing other variables. However, this L2 normal also makes the circuit more complex.
\item We uses Lagrange relaxation via a multiplier $w_i$ to include constraints into objective
\item With the objective function, we turn it into QUBO function $f({\bf x}) = {\bf x}^TQ{\bf x} + {\bf B}{\bf x}^T$, from which we can construct corresponding Hamiltonian: 
\begin{eqnarray}
H_o&=& \sum_{ij}Q_{ij}\frac{(\mathbb{I}+Z)_i}{2}\frac{(\mathbb{I}+Z)_j}{2} + \sum_iB_i\frac{(\mathbb{I}+Z)_i}{2} \\
\quad&=&\sum_{ij}\frac{1}{4}Q_{ij}Z_iZ_j + \frac{1}{2}\sum_i\left(\sum_jQ_{ij}+B_i\right)Z_i + \left(\frac{1}{4}\sum_{ij}Q_{ij}+\frac{1}{2}\sum_iB_i\right)\mathbb{I}\\
\quad&=&\sum_{ij}\frac{1}{2}Q_{ij}Z_iZ_j + \frac{1}{2}\sum_i\left(\sum_jQ_{ij}+B_i\right)Z_i + \left(\frac{1}{4}\sum_{ij}Q_{ij}+\frac{1}{2}\sum_iB_i+\frac{1}{4}\sum_iQ_{ii}\right)\mathbb{I}\\
\end{eqnarray}
\end{itemize}
\item For each constrain that is solved in Dephasing, its Hamiltonian is $H_{d,j}=g_j(x)$ with corresponding unitary $U_{d,j}=e^{i\gamma penalty_j(z)}$. We denote the set of all constraints that are solved via Quantum Penalty Dephasing as $C_d$. In more detail, we have:

\begin{eqnarray}
U_d|z> &=&\prod_{j \in C_d}e^{-i\theta penalty_j(z)}|z>\\
&=&e^{-i \theta\sum_{j \in C_d}(-\alpha)(cost_j-c_j)1_{(cost_j(z)>c_j)}}|z>
\end{eqnarray}
where 
\begin{eqnarray*}
penalty_i(z)=-\alpha(cost(z)-c_i)1_{(cost(z)>c_i)}
\end{eqnarray*}
\item The constraints on ZENO contains a non-unitary circuit to calculate loss and a mixer. We denote the set of all constraints that are solved via Quantum ZENO effect as $C_z$. 



\begin{itemize}
 \item For each constraint $j$, the ZENO Hamiltonian is in the form of subblock matrix, which imply the incomplete ZENO measurements reduce dimension of the state space to $K$ blocks . 
 \begin{eqnarray}
 H_{z,j}=P_jB_j=\sum_{k=1}^{K} P_{k,j}H_jP_{k,j}
 \end{eqnarray}
 \item For each constraint, The ZENO Unitary is then specified as follows, assuming it is measured for $N$ times
\begin{eqnarray}U_{z,j}(z)|z>&=&(Pe^{-iH_jz/N})^N|z>
\\&=&e^{-iP_jH_jz}P_j|z>
\end{eqnarray}
 Frequent measurement converges to a block diagonal in the state space as in \cite{herman2022portfolio}.

 \item With multiple constrains, the aggregate Unitary is a product of individual constraints as 
 \begin{eqnarray}
U_{z}(z)|z>&=&\prod_{j\in C_z}H_{z,j}(z)|z>\\
&=&\prod_{j\in C_z}(Pe^{-iH_jz/N})^N|z>\\
&=&\prod_{j\in C_z} e^{-iP_jH_jz}P_j|z>\\
&=&e^{\sum_{j\in C_z} -iP_jH_jz}\prod_{j\in C_z}P_j|z>\\
 \end{eqnarray}
\item Putting all constraints together, we have the following

\end{itemize}


\end{itemize}
Then the problem after merging the three parts are represented as 
\begin{eqnarray}U(z)|z>&=&U_o(z)*U_z(z)*U_d(z)|z>\\
&=&e^{-i\theta (f(x)+\sum_{j \in C_q}\lambda_j (g_j(x)-\delta_j)^2)}e^{-i \theta\sum_{j \in C_d}(-\alpha)(cost_j-c_j)1_{(cost_j(z)>c_j)}}e^{\sum_{j\in C_z} -iP_jH_jz}\prod_{j\in C_z}P_j|z>
\end{eqnarray}
Please note $H$ can be repeated for a few times when conducting the quantum dephasing operation.

 The corresponding algorithm structure is designed as in Algorithm \ref{alg:hybridframework}. Please note the number of measurement $q$ and the repeating Hamiltonian $p$ are the hyper parameters to set per specific problem. Rational of specific layers of the circuits are discussed in next few sections.

 \begin{algorithm}
        \caption{Hybrid Quantum Circuit integrating QAOA, Dephasing and ZENO}
     \label{alg:hybridframework}
\begin{itemize}
\item Set suitable initial conditions
\begin{itemize}
\item Keep initial solution feasible for all constraints that are subject to quantum ZENO. This can be achieved by pre-run a circuit with constraints in $C_z$.
\end{itemize}
\item Conduct gradient search for $H_o$, for each step:
\begin{itemize}  
\item Conduct dephasing operation for $H_d$ to trigger state transition. The following shrink are accounted for $P$ times where $p=1...P$: 
\begin{eqnarray}
H_{d,p}=\beta_t H_{d,0}+\gamma_p H_{d,f}
\end{eqnarray}
\item Conduct Zeno observation for $q$ times to keep $H_z$ in initial state, i.e. $P_{\boldsymbol{0,z}}(t)$ converges to 1.
 
 Define: 
\begin{eqnarray}
\Theta_{\boldsymbol{H_z}} &=& \langle \psi_{\boldsymbol{0}}| H_z^2|\psi_{\boldsymbol{0}}\rangle - \langle \psi_{\boldsymbol{0}}| H_z|\psi_{\boldsymbol{0}}\rangle ^2
\end{eqnarray} 
\begin{eqnarray} P_{\boldsymbol{0,z}}(t)&=&(P_{\boldsymbol{0}}(\varepsilon))^N \\&=&[1-\frac{\varepsilon^2}{\hbar^2} (\Theta_{\boldsymbol{H_z}})^2 + O(\varepsilon^3) ^N \\
\overrightarrow{large N, small \varepsilon}\quad\ 1-\frac{N*\varepsilon^2}{\hbar^2} (\Theta_{\boldsymbol{H_z}})^2
 &=&1-\frac{t*\varepsilon}{\hbar^2} (\Theta_{\boldsymbol{H_z}})^2
\end{eqnarray}
\item repeat $U(z)|z>=U_o(z)*U_z(z)*U_d(z)|z>$ for $P$ times where $p=1...P$ as objective.

\end{itemize}


\item Exit when the changes of Humiliation in two steps are smaller than given threshold
\end{itemize}
\end{algorithm}

\subsection{Penalty Dephasing as a Circuit Layer}

\begin{figure*}[ht]\centering 
	\includegraphics[width=400pt]{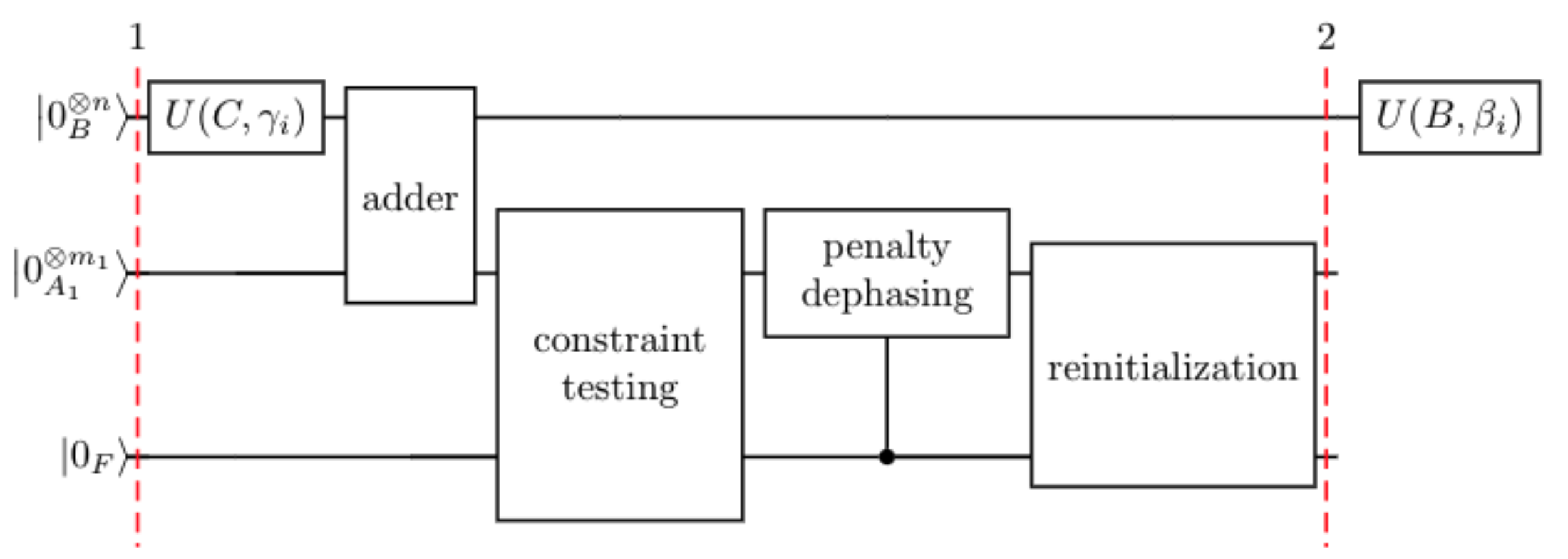}
	\caption{Circuit Structure: Penalty Dephasing Layer}
	\label{fig:dephasingstructure}
\end{figure*}

The structure of penalty dephasing is discussed more in the literature: A cost adder is used to calculate the current value of the constraints based on current states and then the value is rflected into a flag which symbolized the violation of the constraint. With this flag, one further conduct dephasing operation to find new state with lower energy. Therefore, the structure to represent penalty dephasing is designed as in Figure \ref{fig:dephasingstructure}. Please note the demonstration is for a sample problem with one constraint and the phase return of a problem with multiple constraints are integrated in front of all constraints. 

Each block is explained in more debail below:
\begin{itemize}
    \item $U(C,\gamma)$ is the phase return part, which can be the obj without constraints or with the remaining constraints (that are not covered by zeno or dephasing) which are modeled within inner layers
    \item $adder$ is used to calculate the constraint value for a given solution in circuit.
    \item $constraint testing$ compares the sum with the threshold.
    \item $penalty dephasing$ conducts quantum dephasing and enforce the constraints.
    \item $reintialization$ restores the mixer for every iteration.
 \end{itemize}

Overall the calculation block can repeat $P$ times. For simplicity, $P=1$ is selected in numerical example

\subsection{Quantum ZENO Effect as a Circuit Layer}

\begin{figure*}[ht]\centering 
	\includegraphics[width=400pt]{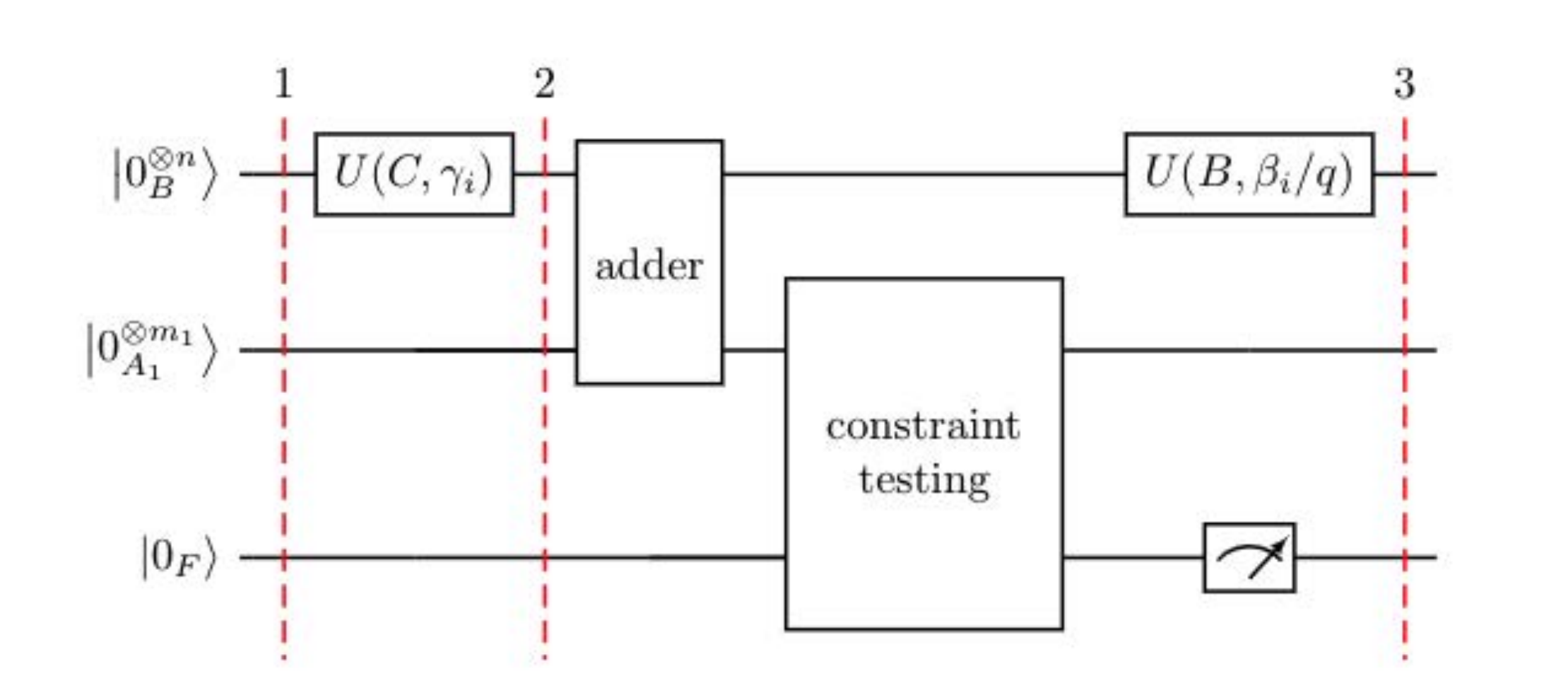}
	\caption{Circuit Structure: ZENO layer}
	\label{fig:zenostructure}
\end{figure*}

The structure for introducing the quantum ZENO effect to represent a constraint has not been established in previous literature. The intuition is to leverage some existing blocks in panalty dephasing, including cost adder and constraint testing. However, we replaced the penalty dephasing and re-initialization with frequent measurements and let ZENO to maintain the status of the flag corresponding to the selected constraint. 
Particularly the structure to represent Quantum ZENO effect is designed as in Figure \ref{fig:zenostructure}. Please note the demonstration is for a sample problem with one constraint and the phase return of a problem with multiple constraints are integrated in front of all constraints.

Each block is explained in more detail below:
\begin{itemize}
    \item $U(C,\gamma)$ is the phase return part,which can be the obj without constraints or with the remaining constraints (that are not covered by zeno or dephasing) which are modeled within inner layers
    \item $adder$ is used to calculate the constraint value for a given solution in circuit.
    \item $constraint testing$ compares the sum with the threshold and set a flag
    \item A measurement is conducted to the flag and enforce the constraints. Please note several measurements can be conducted within one given cycle of mixer
    \item the mixer part is also separate to $Q$ sub blocks to align with the $Q$ measurement
 \end{itemize}

To maintain the constraint in the proper state, it is best practice to start state evolution based on the superposition of all feasible states corresponding to the constraints that are solved by ZENO. Then a two step procedure are proposed:
\begin{itemize}
\item Pre-run circuits with ZENO constraints and post select the feasible states
\item Initialize the full circuit using superposition of selected states and obtain the final results.
\end{itemize}

Overall the calculation block can repeat $P$ times, and during each iteration, during each iteration, $Q$ measurements can be conducted. For simplicity, $P=1$ and $Q=1$ are selected in numerical example.

\subsection{Structure of the Hybrid Circuit}

\begin{figure*}[ht]\centering 
	\includegraphics[width=400pt]{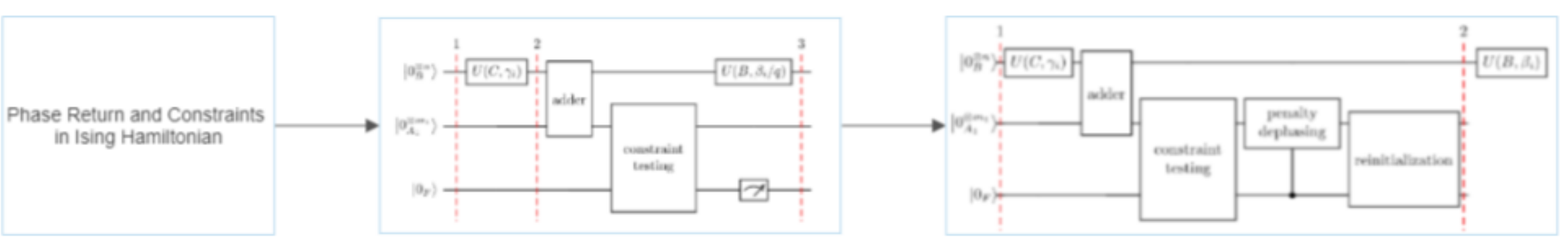}
	\caption{Additive Hybrid Circuit Structure}
	\label{fig:HybridStructure_smartdraw}
\end{figure*}

To reuse the qubit and reduce complexity of the circuit, we take a linear additive structure to combine the circuit of QAOA with that of Dephasing and Zeno. Different layers are attached to the right of the basic circuit for phase return, as in Figure \ref{fig:HybridStructure_smartdraw}. We try to put the circuit of the same type together and adjust the same initial state, Such as block-based structure makes the framework scale-able and reliable.

\section{ Properties of Quantum Circuit as an Optimization Routine}

In this section, some properties are discussed and tested for such algorithms under a typical binary optimization problem. We introduce the problem and performance measure first and then show the properties of the algorithms respectively.

\subsection{Introduction of Airbus Cargo Loading Problem}

The full description of Airbus cargo loading problem is provided in Airbus website and a mathematical description in algebra form is further discussed in Giovanni(2019). The purpose of this use case is to find the optimal solution of loading maximum number of cargo within the constraints of airplane's capacity, weight it can bear, etc. 

To further simplify the problem for the purpose of our experiment, we started off with the use case as: 3 cargo, 2 positions to load. Three Cargo weights are 1,2,3 (regardless the weight unit), maximum payload weight limit is 3 (regardless the weight unit), all the cargo have same size and each of them can be fit into 1 position.

The air cargo problem can be summarized as following. Let $x_i$ be binary variables ($x_i=0,1$), we aim to

maximize the objective function total weights
\begin{eqnarray}
 {max}\sum_j\sum_i w_ix_{ij}\\
\end{eqnarray}

subject to the weight constraint:

\begin{eqnarray}
\sum_j\sum_iw_ix_{ij}\leq W\\
\end{eqnarray}

subject to the position constraint(one position can only take one cargo):

\begin{eqnarray}
\sum_ix_{ij}\leq 1\quad\\
\end{eqnarray}

subject to the cargo constraint(one cargo can only appear at one position):

\begin{eqnarray}
\sum_jx_{ij}\leq 1\\
\end{eqnarray}

Upon introducing slack variable, the problem is equivalent to minimize the following cost function 

\begin{eqnarray}
f &=& -\sum_j\sum_iw_ix_{ij} + P_w(\sum_j\sum_iw_ix_{ij} + \sum_k2^kC_k - W)^2 \\
&&+P_p\sum_j(\sum_ix_{ij} + \sum_p2^pC_p - 1)^2 + P_c\sum_i(\sum_jx_{ij} + \sum_c2^cC_c - 1)^2
\end{eqnarray}

where $P_w$, $P_p$ and $P_c$ are the Lagrange multipliers.
For simplicity, we fix them to be of the same value $W$ in this paper.

In the numerical tests, we solve a simple case where there are 2 position and 3 cargos with weights 1,2 and 3respectively. Total weight constraint is 3. All cargos are with size 1.

\subsection{Performance and Cost Measures for a Quantum Circuit}
We leverage IBM qiskit as the implemetation for quantum circuits and the performance measure of the circuits include below \cite{QuantumCir_IBM}:
\begin{itemize}
\item Probability of gain feasible solutions. 
\item Probability of gain optimal solutions.
\item Execution Speed on a given CPU
\end{itemize}

Complexity measures for the circuits include below \cite{QuantumCir_IBM}:
\begin{itemize}
\item Number of non local gates: number of non-local gates (i.e.involving 2+ qubits)
\item number of qbits number of qubits
\item number of clbits number of classical bits
\item depths: circuit depth (i.e., length of critical path)
\item width: number of qubits plus clbits in circuit
\item size: total number of gate operations in circuit.
\item num of parameters: the number of parameter objects in the circuit.
\item num of unitary factors: How many non-entangled subcircuits can the circuit be factored to
\end{itemize}

\subsection{Explore the Solution Space More Efficiently}
In most of classical optimization algorithms, a solution path is generated and a new solution is explored at each step. An exception is genetic algorithm, a set of existing solutions can be muted or crossed to generate a set of new solutions. 

\begin{figure}[htb]
  \begin{center}
   {\includegraphics[angle =0 , width=0.48\textwidth]{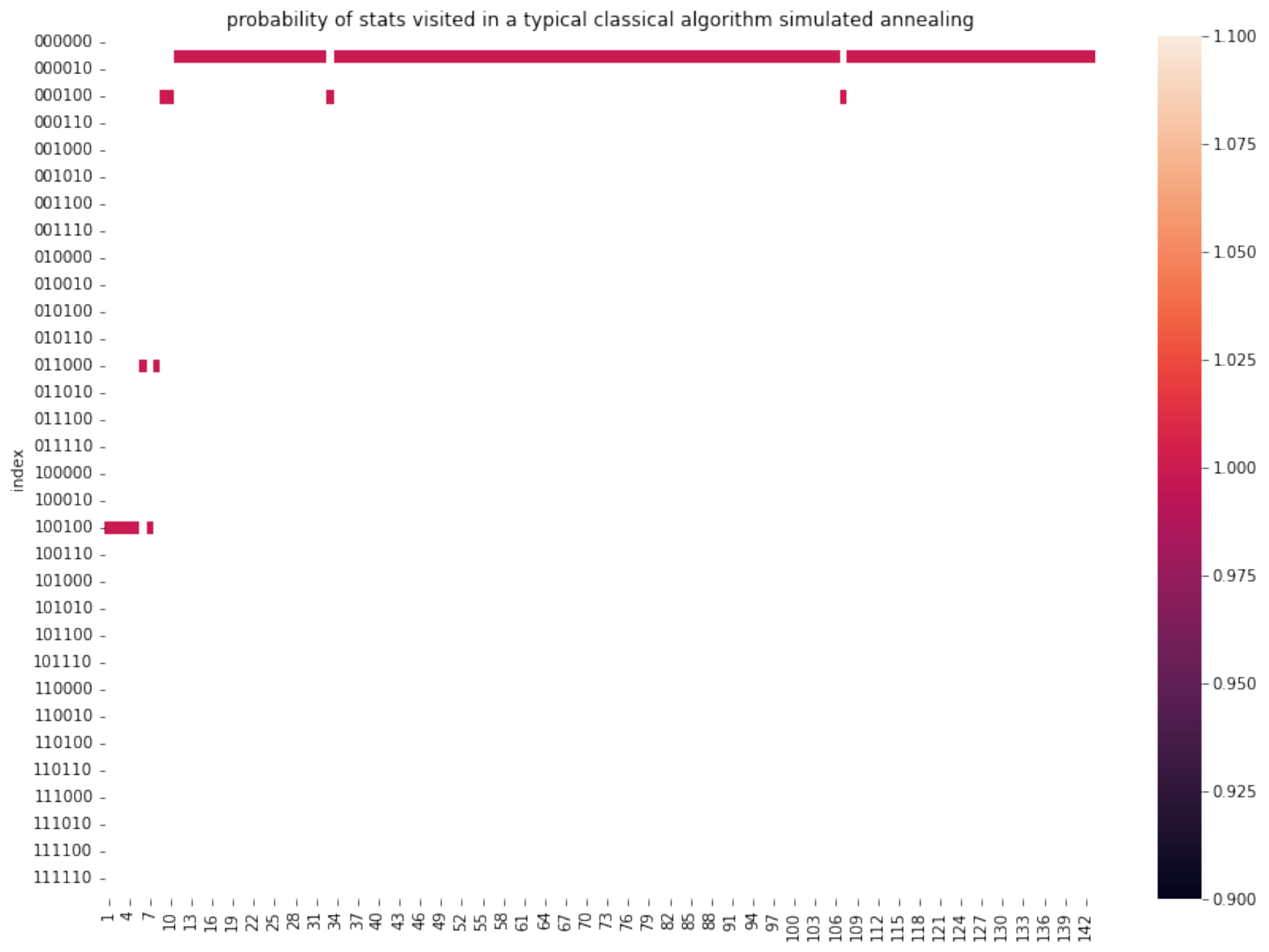}}
    {\includegraphics[angle=0 , width=0.48
  \textwidth]{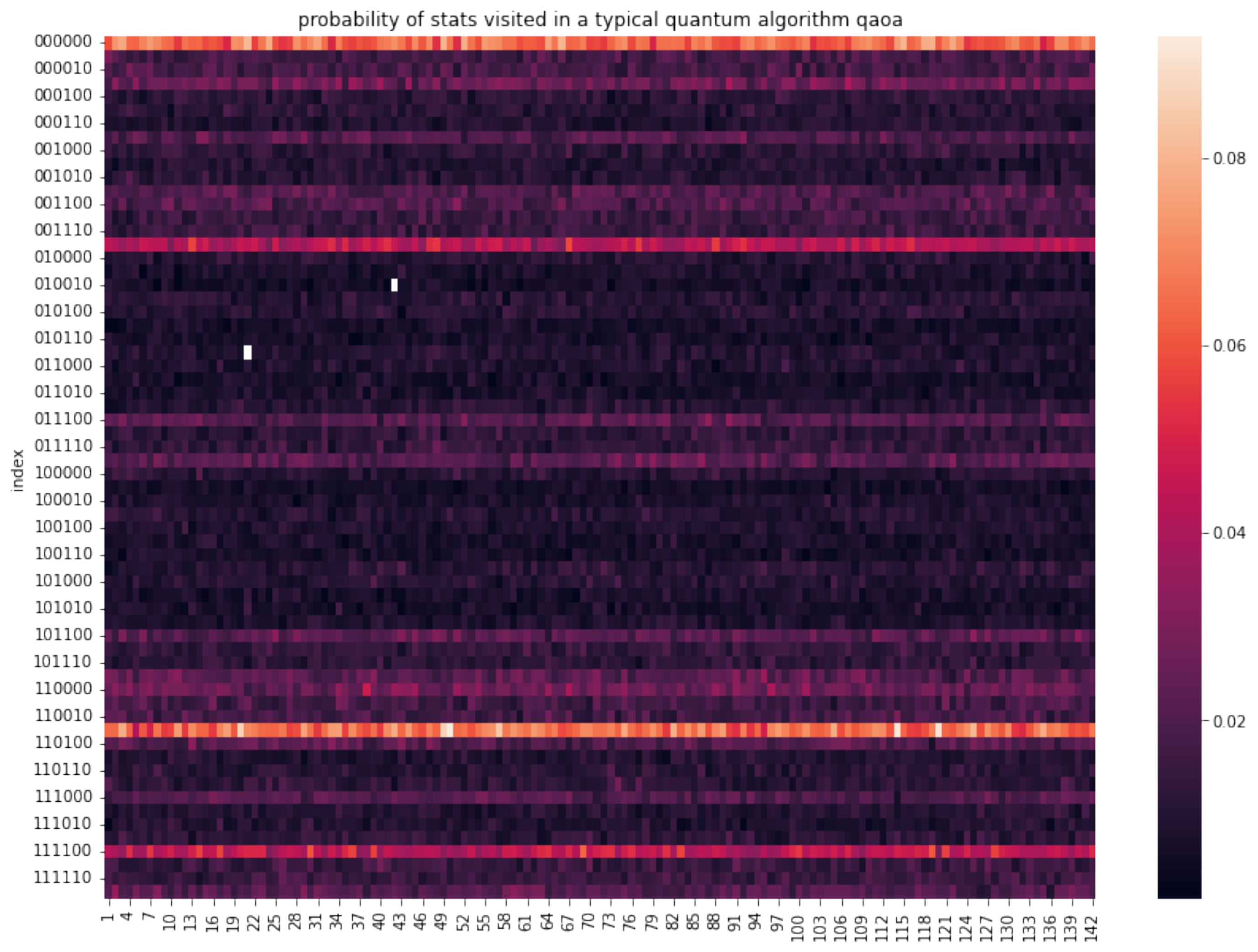}}
  
  {\includegraphics[angle=0 , width=0.48
  \textwidth]{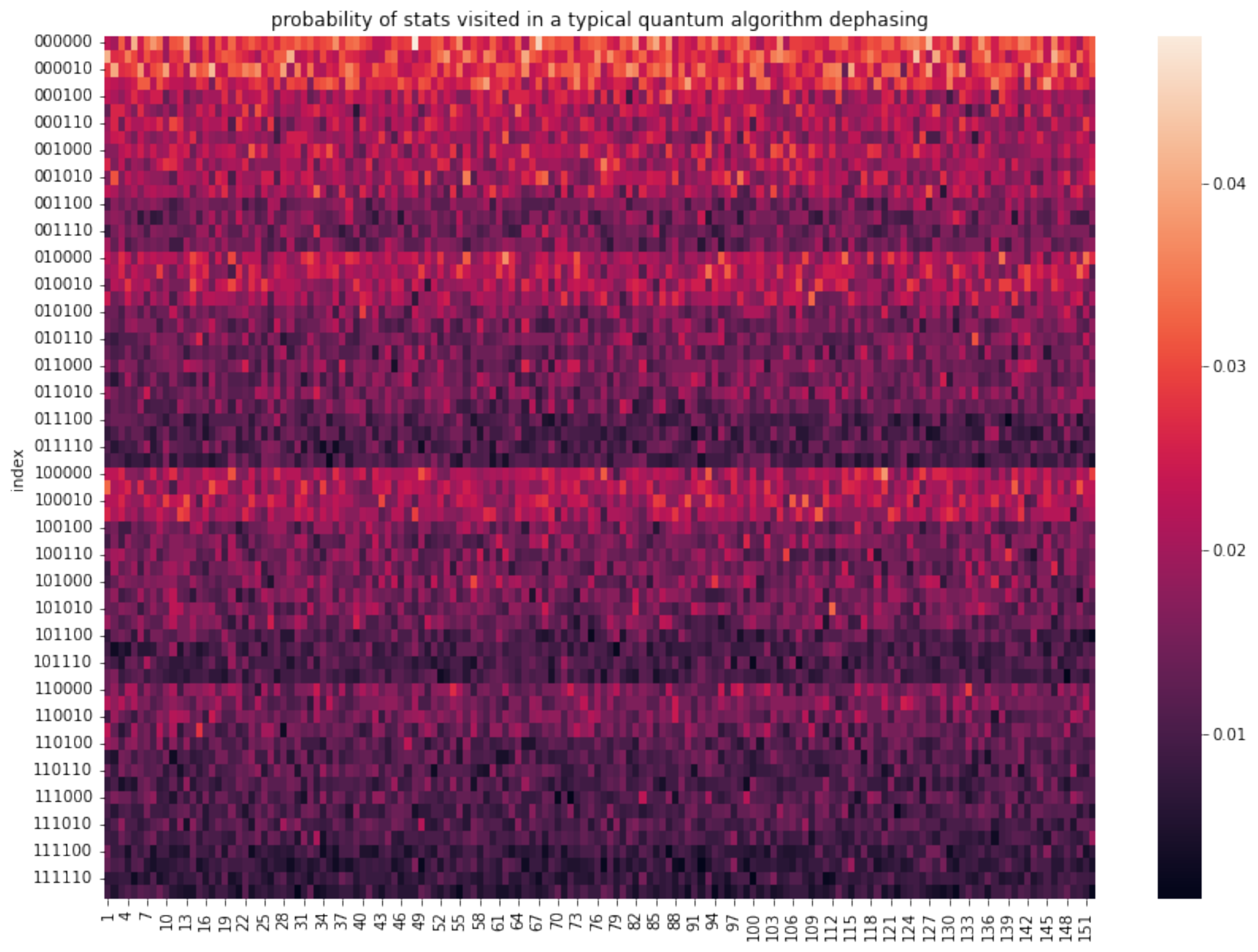}}
  {\includegraphics[angle=0 , width=0.48
  \textwidth]{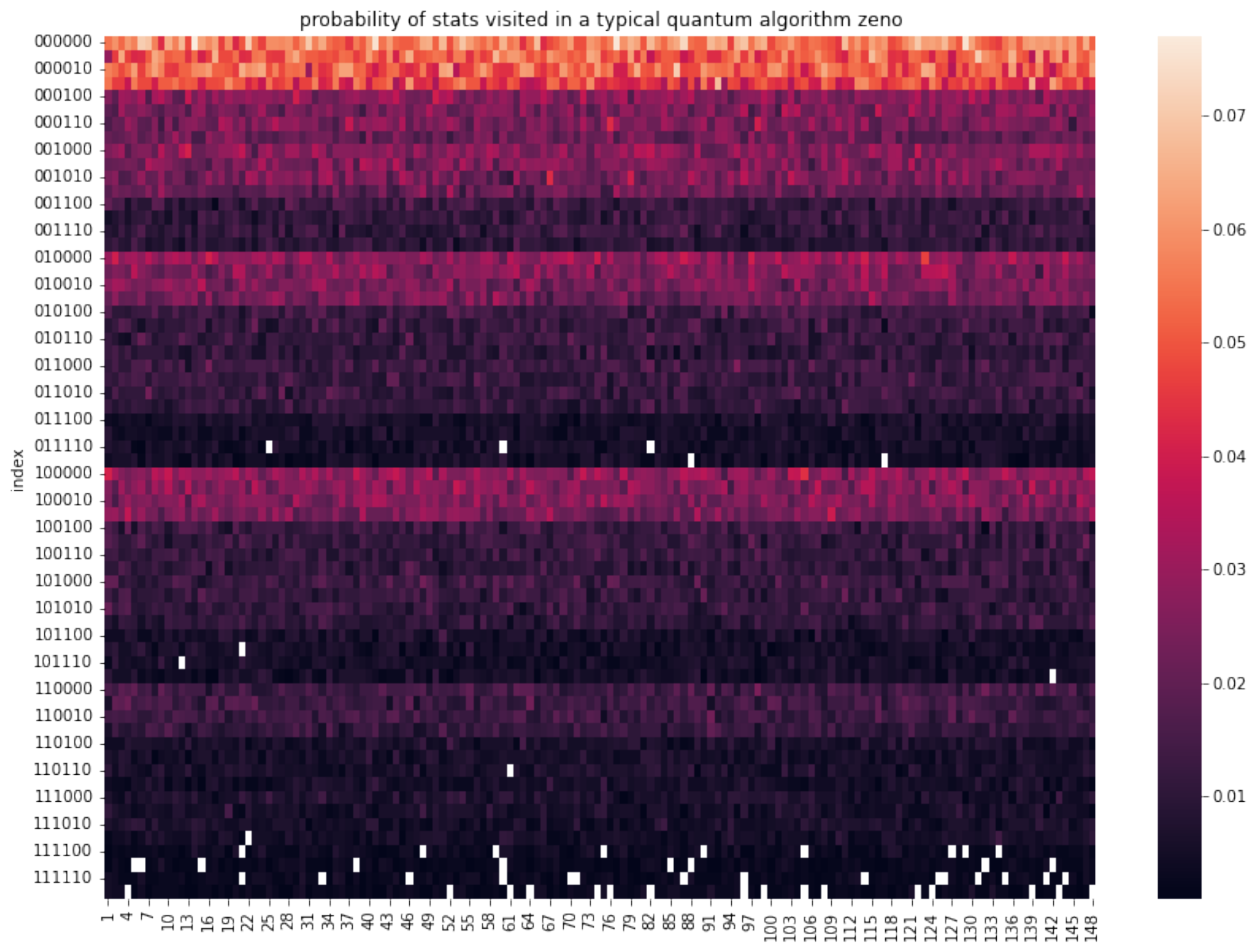}}
  \caption{Comparison of probability of stats visited (classical algorith, qaoa, dephasing and zeno)}
  \label{fig:state-visited}
  \end{center}
  \end{figure}

In the contrast, a quantum algorithm explore the all solution space through superposition; at each step of the algorithm, all solutions have certain probability to show up in measurements. The algorithm just gradually increase the probability of observing the optimal solution via suitable quantum operation such as rotation, dephasing etc. The algorithm stops when the quantum system stopped at a low energy state. Moreover, the initial state setting may depend on specific the theoretical requirements of quantum algorithms involved, 
For test purpose, we compared the three circuits where all constraints are solved in QAOA, or all in Dephasing v.s. all in ZENO. These are the three most extreme cases in the family of hybrid circuits, 
As shown in Figure \ref{fig:state-visited}, all quantum algorithms covered the spectrum of solutions while the classical algorithm (simulated annealing) only explore a single solution at each step. Among the quantum algorithms, their pattern differ due to differences in rotating phases.

The advantages of superposition is that one can explore the solution space of a complex optimization problem with a single run of quantum algorithms. Efficient frontier which shows the changes of solution due to changes of Lagrange multipliers can be identified more efficiently, as demonstrated in next few sections.

\subsection{Changes of Cost Function and Feasible Probability during the Iteration}

\begin{figure}[htb]
  \begin{center}
   {\includegraphics[angle =0 , width=0.48\textwidth]{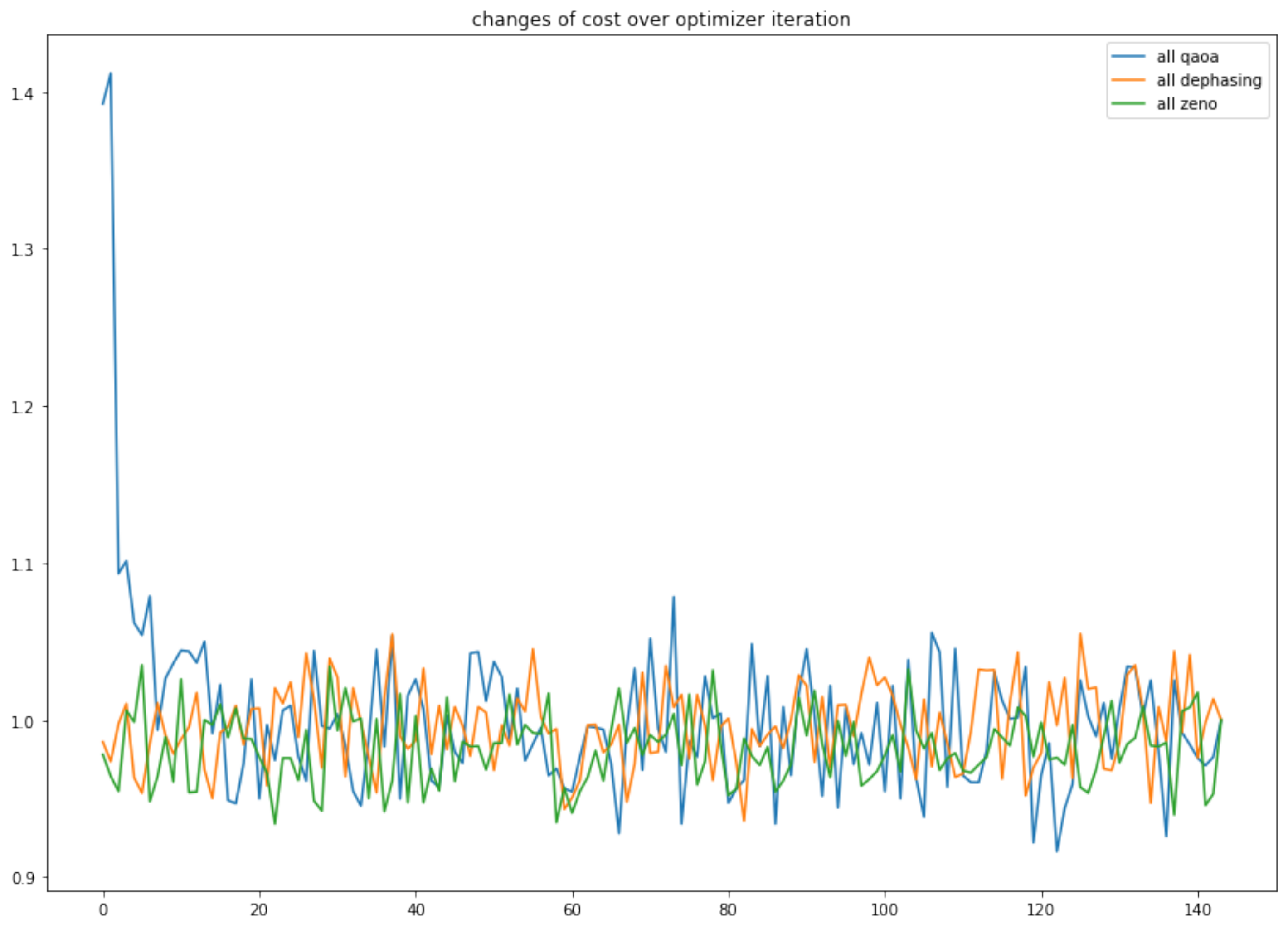}}
    {\includegraphics[angle=0 , width=0.48
  \textwidth]{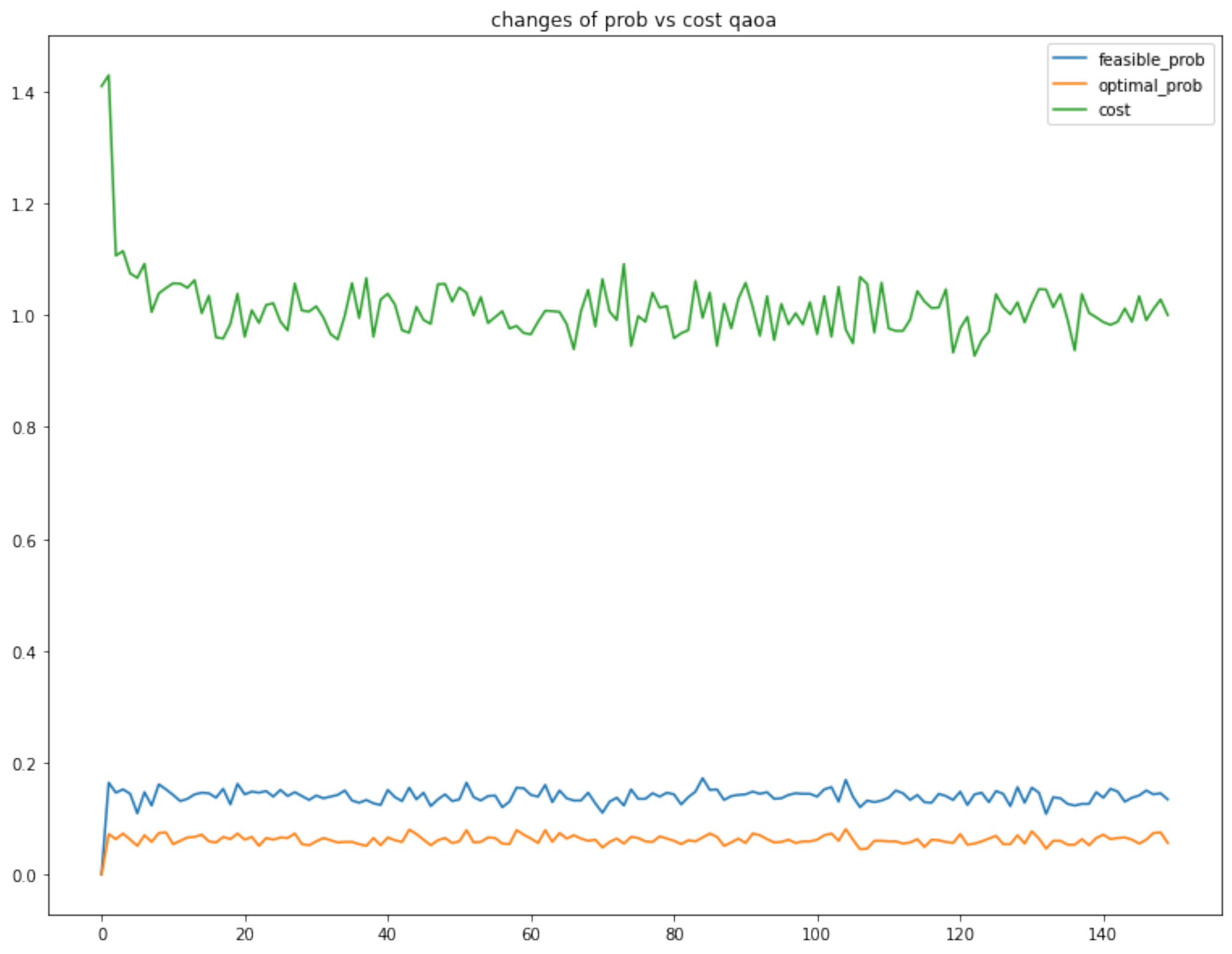}}
  
  {\includegraphics[angle=0 , width=0.48
  \textwidth]{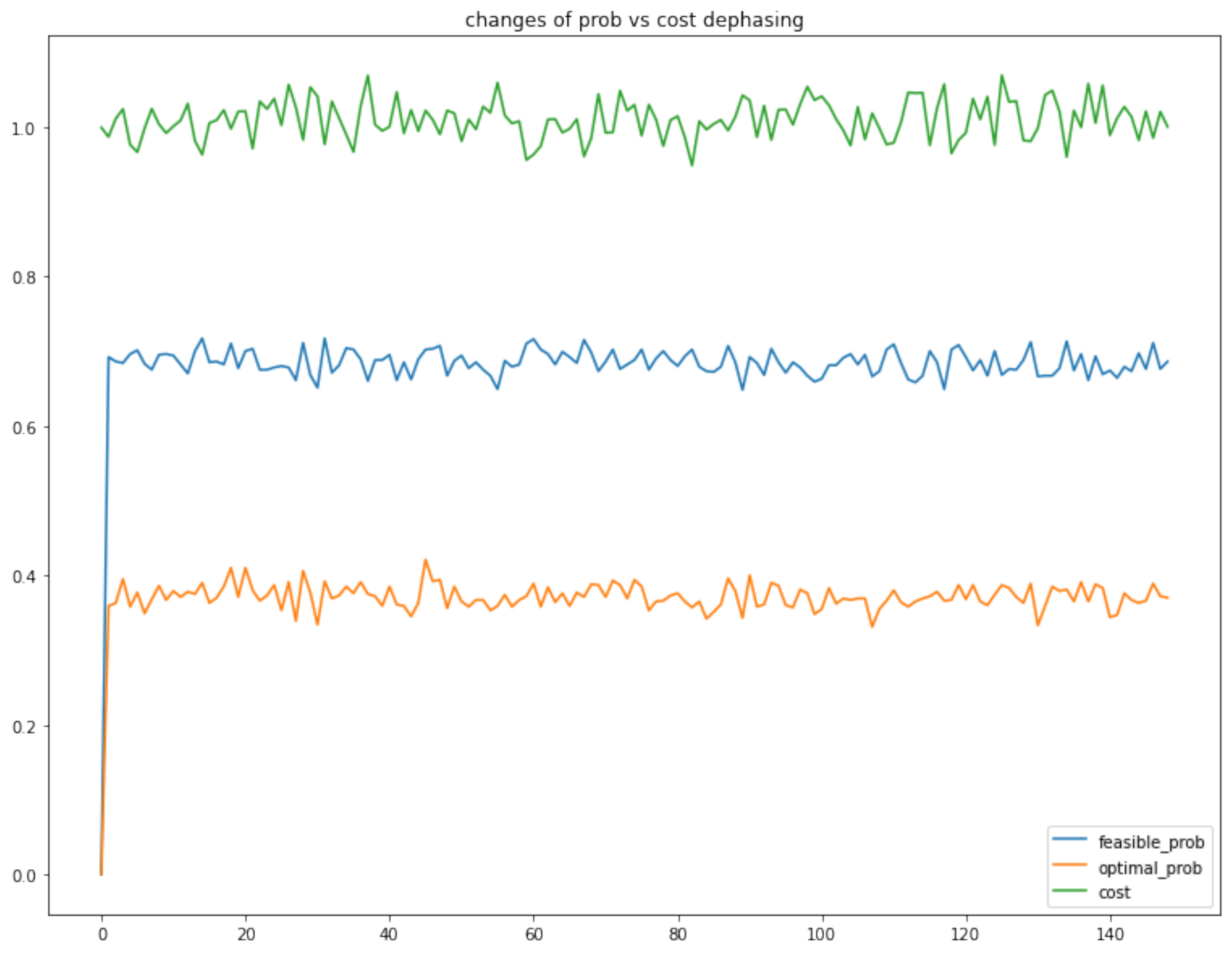}}
  {\includegraphics[angle=0 , width=0.48
  \textwidth]{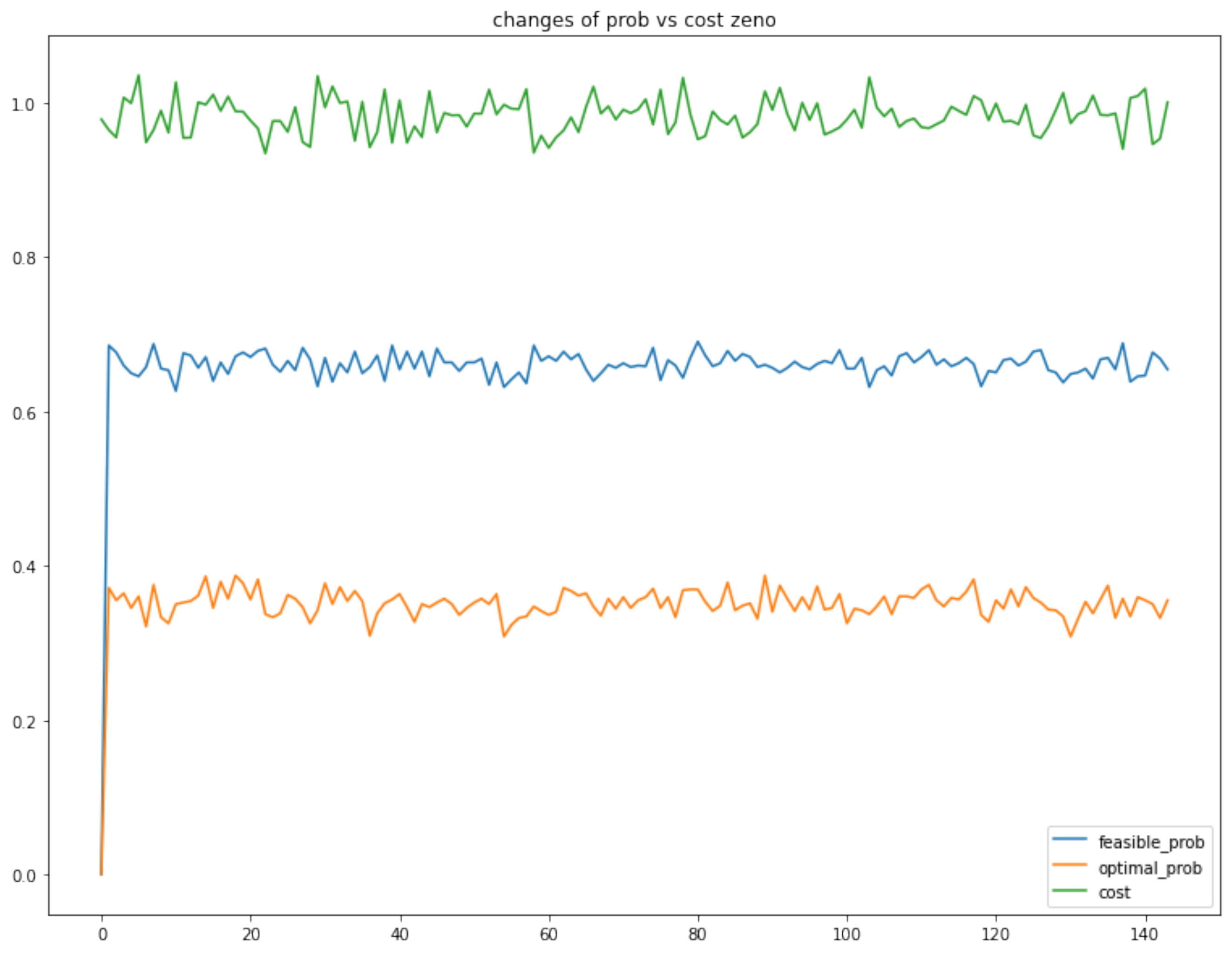}}
  \caption{Comparison of cost and probability of obtaining feasible/optimal solutions over iteration (qaoa, dephasing and zeno)}
  \label{fig:costandprob}
  \end{center}
  \end{figure}

We  extracted the changes of cost function and probability of obtaining feasible/optimal solutions during each step of the numerical optimization routine. For simplicity, we still test the three circuits introduced in the last section.
As in the first figure in Figure \ref{fig:costandprob}, it is observed that 
\begin{itemize}
\item QAOA by nature required numerical optimization to identify lower system states hence the cost function drops significantly over each iteration
\item ZENO or dephasing, lower state in their structure hence the state is lowered in the first iteration resulting in now significant drop. We keep ZENO and dephasing in numerical optimization routine just to keep the same structural framework for the whole family.
\item If some constraints are solved in QAOA while others are solved in ZENO and dephasing, we choose to run whole circuits through the numerical optimization routine to minimize the objective.
\end{itemize}


In the last three figures in Figure \ref{fig:costandprob}, it is further observed that there are negative correlations between changes of cost function and feasible (optimal) probability during the optimization. The lower cost function value is, the more likely it is to identify feasible (optimal) solutions. Please note there exist an increase of probability for Dephasing and ZENO during the first loop, which shows that system state is evaluated and lowered due to their own specific nature (not depending on the numerical optimization).  

\subsection{Sensitivity test of Lagrange Multipliers}

\begin{figure*}[ht]\centering 
	\includegraphics[width=300pt]{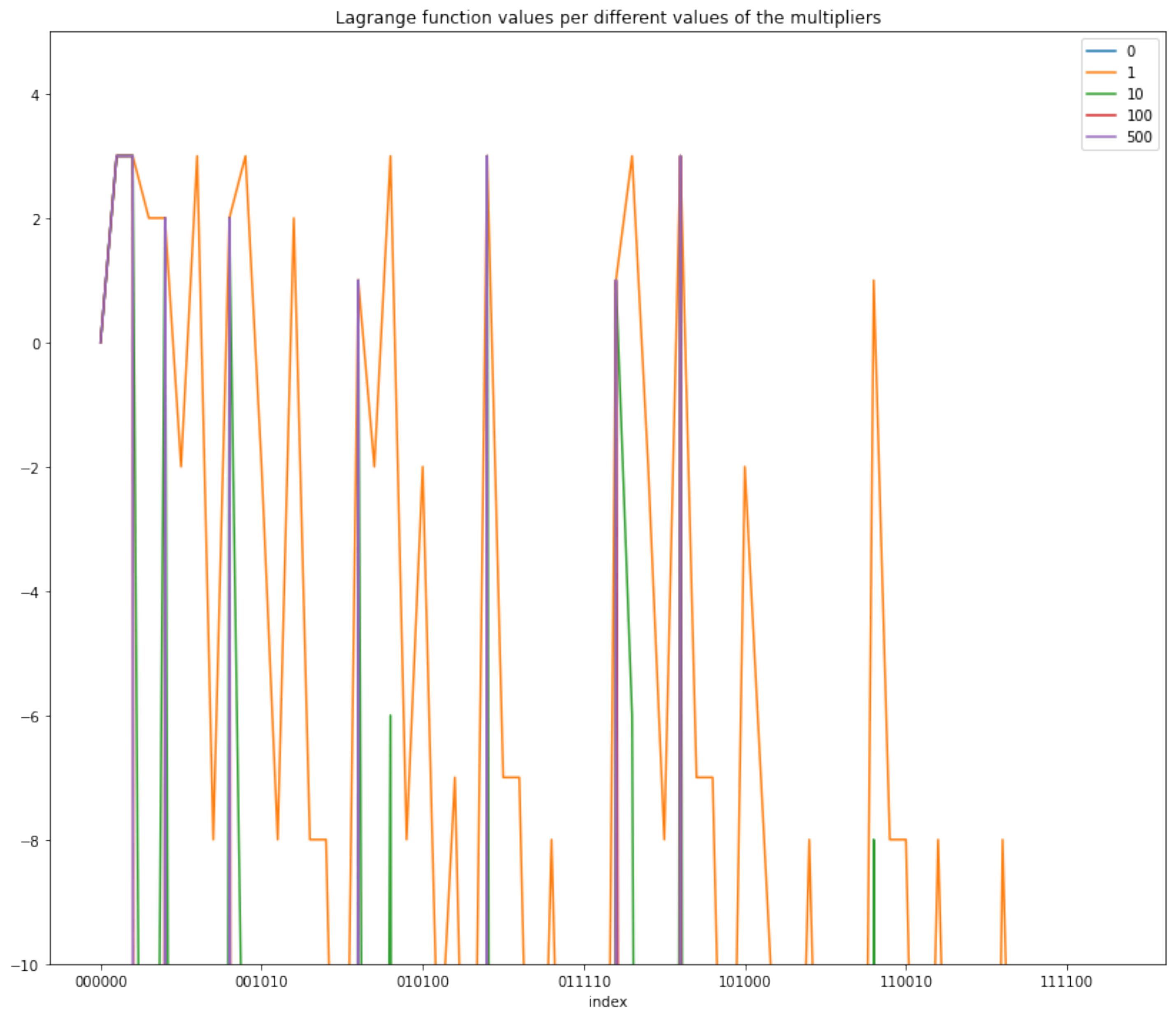}
	\caption{Impact of Lagrange multiplier to objective value}
	\label{fig:LM_impact_on_obj}
\end{figure*}

This section further access the impact of Lagrange Multipliers on optimal solution and probability of obtaining optimal solution,

In Figure \ref{fig:LM_impact_on_obj}, we plotted the changes of Lagrange function values for the sample problem per different values of Lagrange multipliers. The peaks are the optimal value while only the blue peaks are the optimal real objective function within the feasible region. Given all solutions, one can generate such figures. What makes difference here, one just has to run a quantum algorithm once and it will show the spectrum of solutions so that such analysis can be done. If one uses classical optimization routine, he has to tweak the values of Lagrange multiplier and resolve the problem multiple times. Similarly, in quantitative finance, efficient frontiers can be generated more efficiently using a quantum algorithm.

\begin{figure*}[ht]\centering 
	\includegraphics[width=300pt]{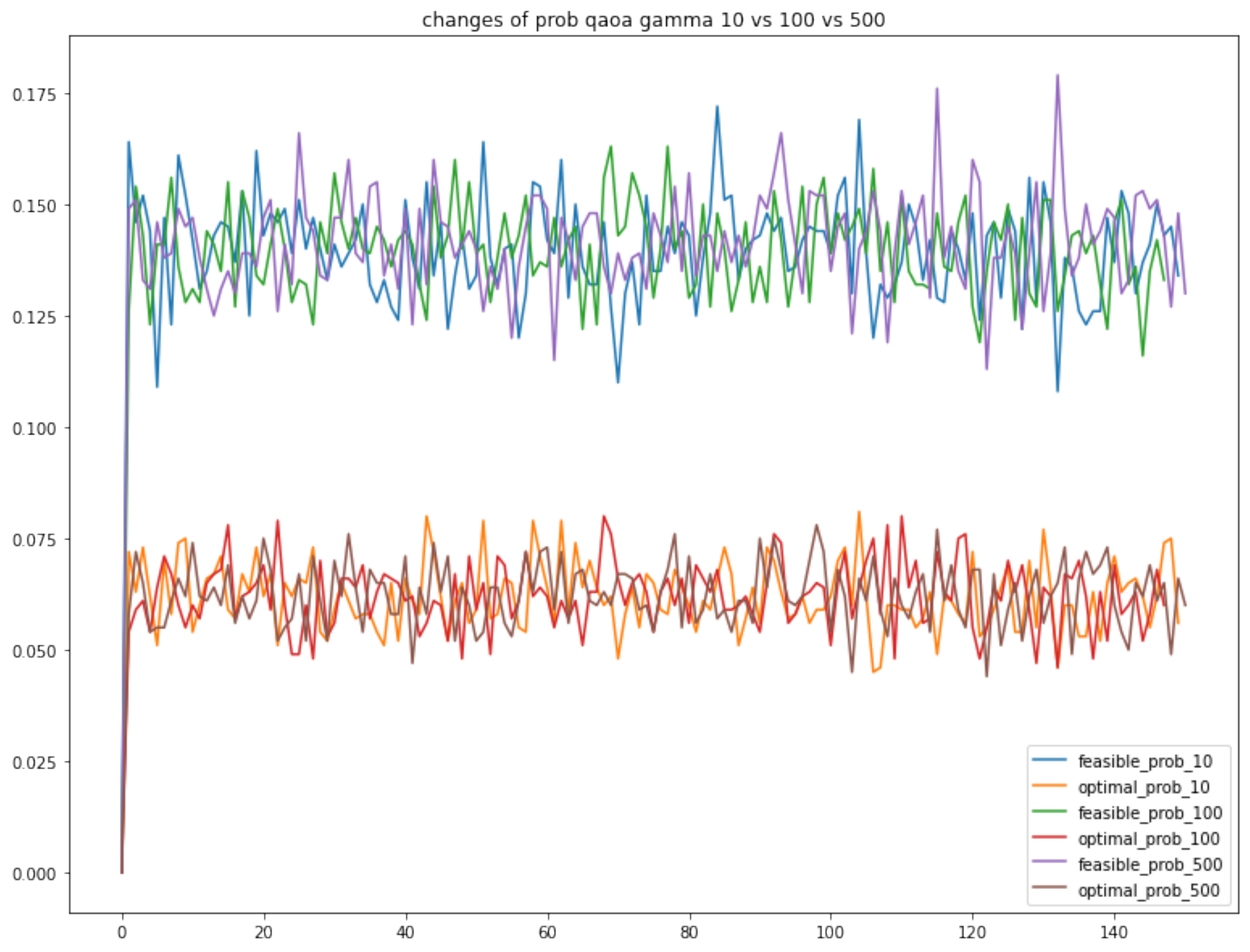}
	\caption{Impact of Lagrange multiplier to probability of obtaining feasible or optimal solutions}
	\label{fig:LM_impact_on_RPOB}
\end{figure*}

In Figure \ref{fig:LM_impact_on_RPOB}, we also tweaked the Lagrange Multiplier and checked the changes of the probability of obtaining the feasible and optimal solutions. Intuitively, the higher the multipliers values are , the cost of being infeasible is greater, the higher probability one should expect to find feasible solutions, and it is consistent with the observation. Meantime, the probability of obtaining the optimal solution also increases as Lagrange multiplier value increases.


\section{Performance of the Hybrid Circuits}

In this section, we further test the performance of the hybrid circuits.

\subsection{Different Representation of the Weight Constraint}
As it is mentioned above, the maximum mass constraint contains a big integer number which consumes quite a few qubits to encode in standard Ising form. And the number of gates increases as $O(n^2)$ in standard Ising circuit. To simplify the circuit, we first treat this constraint in Zeno or Penalty dephasing,while keeping other constraints under standard QAOA.

Essentially, the three combination are tested initially

\begin{itemize}
\item(6,0,0) all constraints in QAOA representation 
\item(5,1,0) Only keep weight constraint in Zeno representation 
\item(5,0,1) Only keep weight constraint in dephasing representation
\end{itemize}

\begin{figure*}[ht]\centering 
	\includegraphics[width=300pt]{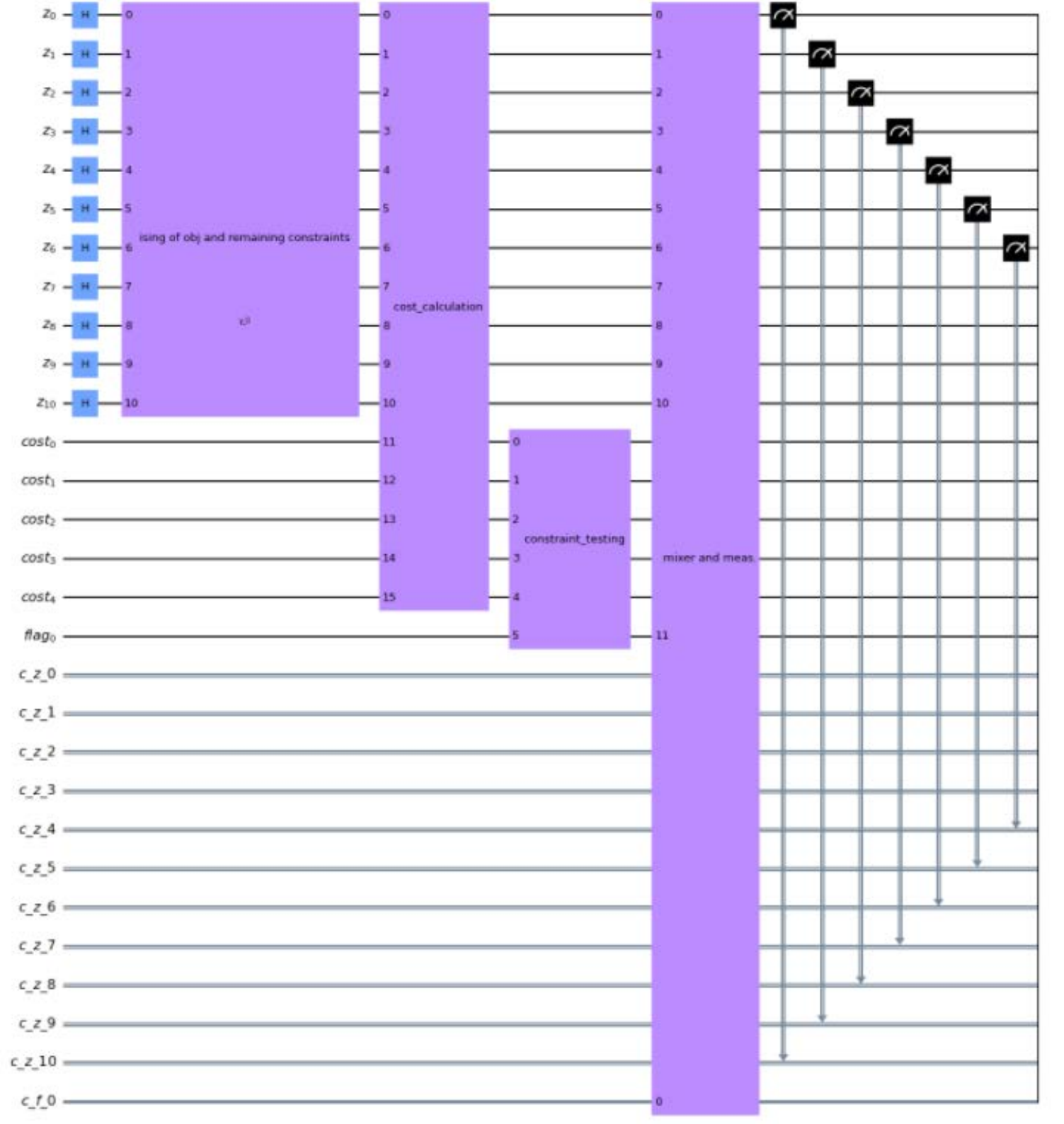}
	\caption{Circuit Structure: QAOA with Zeno}
	\label{fig:QAOAwithZENO}
\end{figure*}

The structure of the QAOA with Zeno is displayed in Figure \ref{fig:QAOAwithZENO}, where target function and constraints that are in standard Ising form are treated as phase return in the circuit.

\begin{figure*}[ht]\centering 
	\includegraphics[width=400pt]{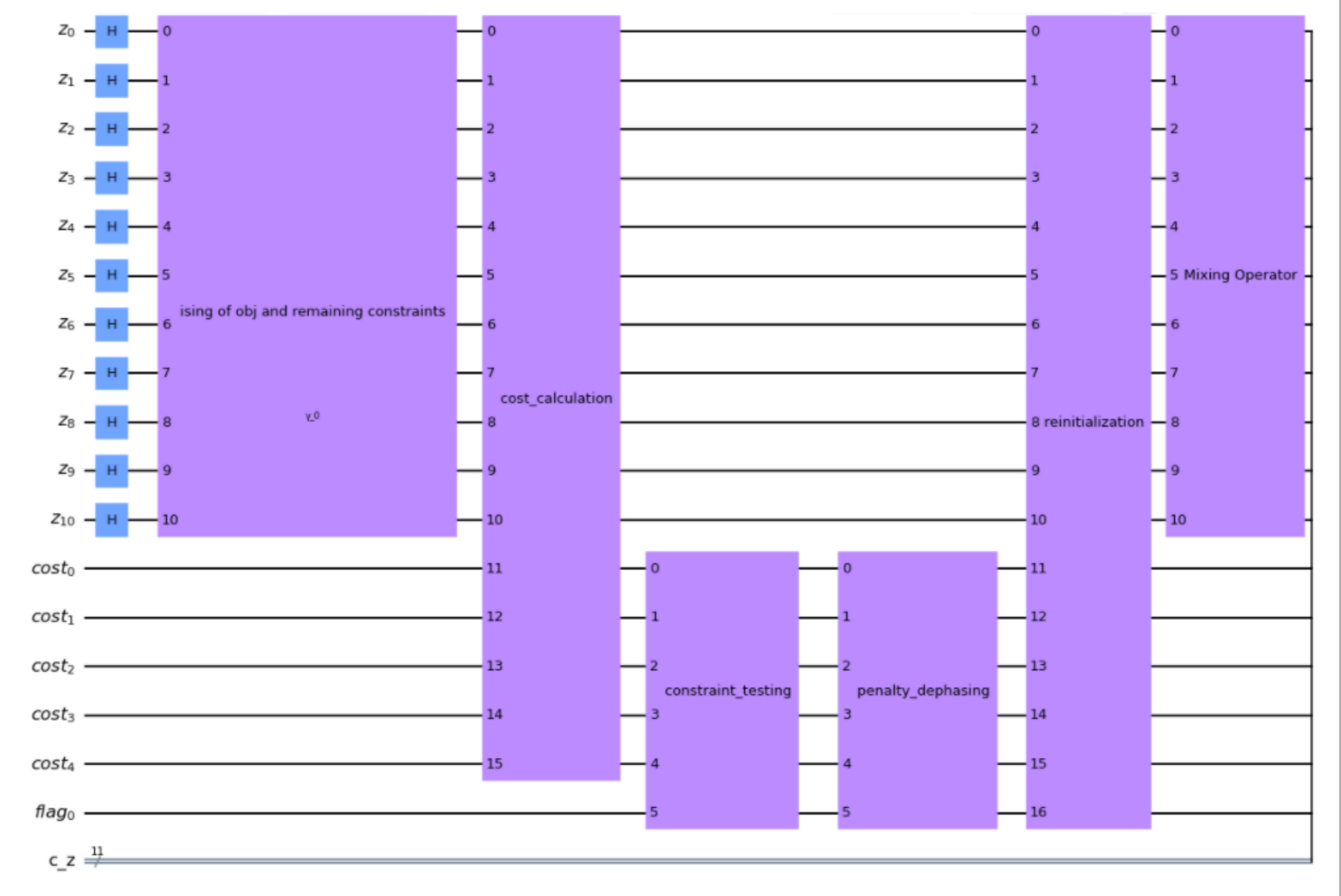}
	\caption{Circuit Structure: QAOA with Penalty Dephasing}
	\label{fig:QAOAwithDephasing}
\end{figure*}

The structure of the QAOA with penalty dephasing is displayed in Figure \ref{fig:QAOAwithDephasing}, where  target function and constraints that are in standard Ising form is treated as return in the circuit.

\begin{figure*}[ht]\centering 
	\includegraphics[width=200pt]{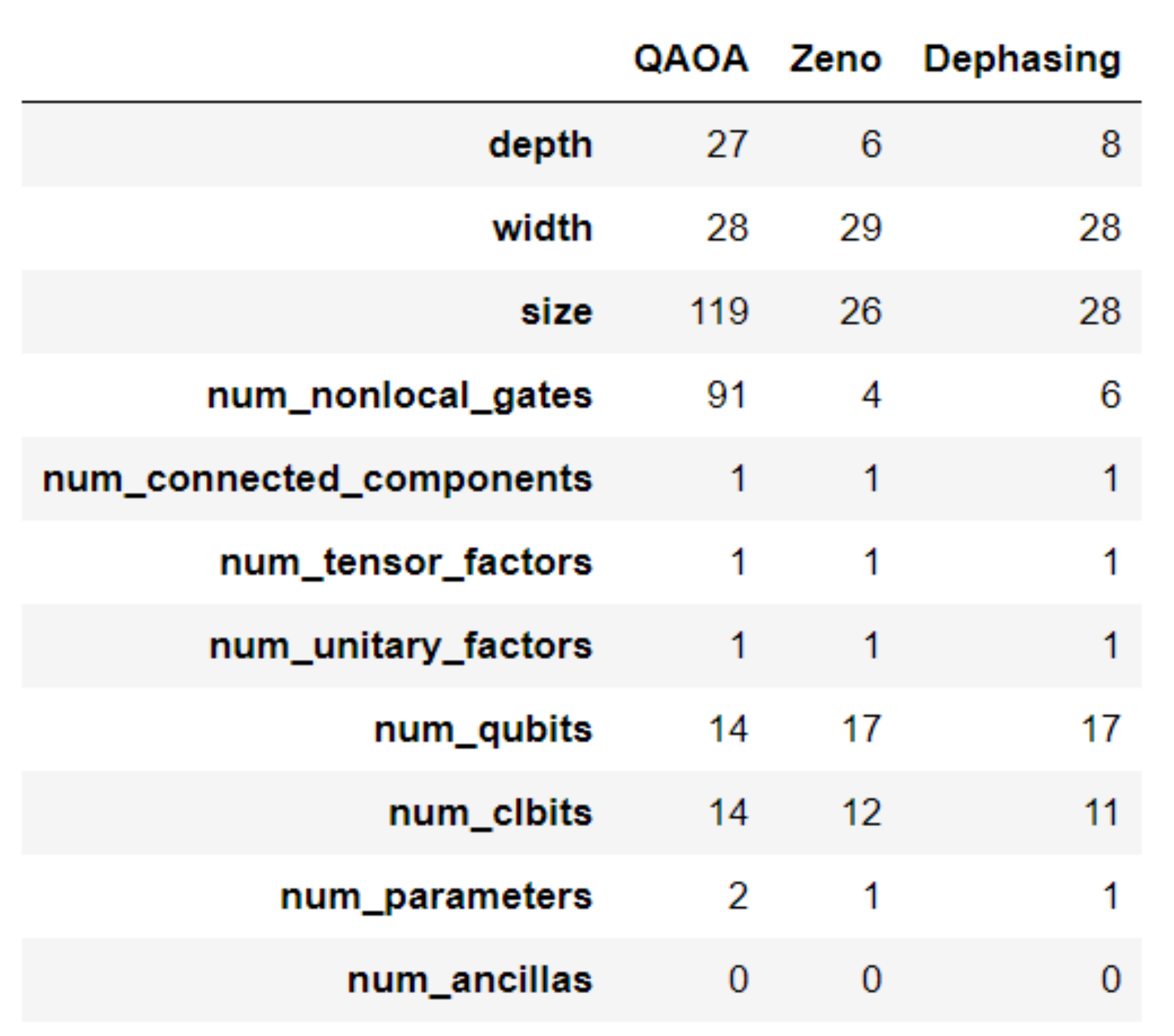}
	\caption{Circuit Structure: QAOA with Zeno}
	\label{fig:CircuitStatsComparison}
\end{figure*}

The basic stats of the three circuits are compared in Figure \ref{fig:CircuitStatsComparison}, please note QAOA with Zeno/penalty dephasing reduces non-local(complex) gates significantly with slightly more qubits.

\begin{figure*}[ht]\centering 
	\includegraphics[width=200pt]{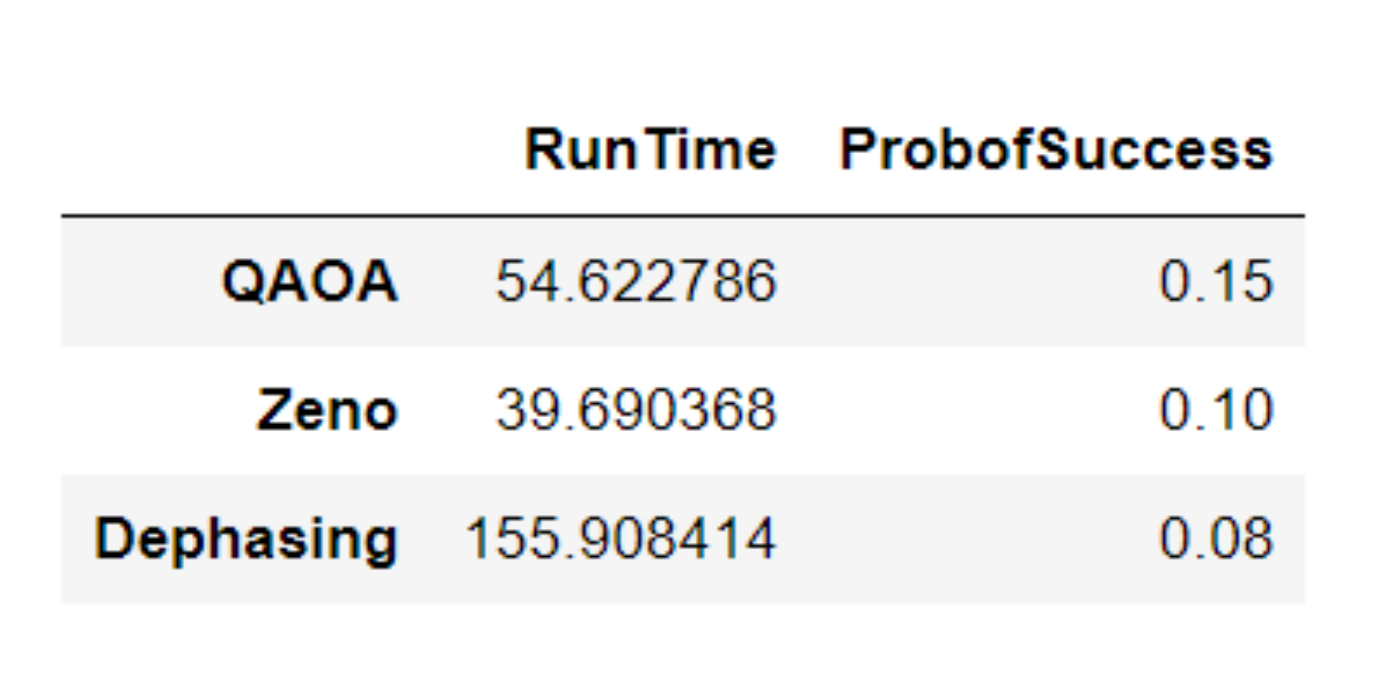}
	\caption{Circuit Structure: QAOA with Zeno}
	\label{fig:CircuitPerformanceComparison}
\end{figure*}

The performance stats of the three circuits are compared in Figure \ref{fig:CircuitPerformanceComparison}, It is observed the probability of finding optimal solution decreases slightly with Zeno or penalty dephasing. 

\subsection{Different Representations of All Constraints}

There are in total 3 type of constraints (6 constraints: 2 position constraint, 3 cargo constraint, 1 weight constraint) in our simplified problem, hence it leads to a family of 729 circuits with slightly different structures. In this section, we implemented all algorithms and show their relative performance measure, to illustrate consistent observation with the theoretical properties:

\begin{itemize}
\item (a) {QAOA applied to all constraints}
Treat all constraints as part of objective function in standard QAOA optimization. This is the most standard quantum opmization methods against such a binary optimization problem.

\item (b) {Dephasing applied to all constraints}

As mentioned in previous section , Penalty Dephasing contains multiple components: Phase Return (which represent the obj function without constraints), Adder(Cost Calculation), Constraint Testing, Penalty Dephasing, Re-initialization. Given the nature of this approach is to only add penalty driven phase to the possible states that didn't pass the constraint testing based on the cost value the adder calculated, and then the qubits for Adder, Constraint Testing will be reset to initial state during Re-initialization,so we will keep Adder, Constraint Testing, Penalty Dephasing and Re-initialization part of the circuit for each constraint (the part of diagram repeat 6 times) and to see whether accumulated penalty dephasing can help us find best solution

\item (c) {Zeno applied to all constraints}

Phase Return (which represent the obj function without constraints) stays the same as previous approach. One standard way of introducing Quantume Zeno Effect is to perform frequent projective measurements. Within each zeno block we constructed for each constraint, we perform project measurement n times. 

\item (d) {QAOA, Dephasing and Zeno applied to different subsets of constraints}

Phase Return (which represent the obj function without constraints) stays the same as previous approach, we use QAOA, Dephasing and Zeno to implement the constraints
\end{itemize}

To be most general, given there are $n$ constraints, each of which are with 3 choice of implementation; standard QAOA, dephasing and ZENO, there are in total $3^n$ circuits that we can generate to solve the same constrained optimization problem. Below we test this case for the problem with 6 constraints, resulting in total 729 different quantum circuits.

\begin{figure*}[ht]\centering 
	\includegraphics[width=400pt]{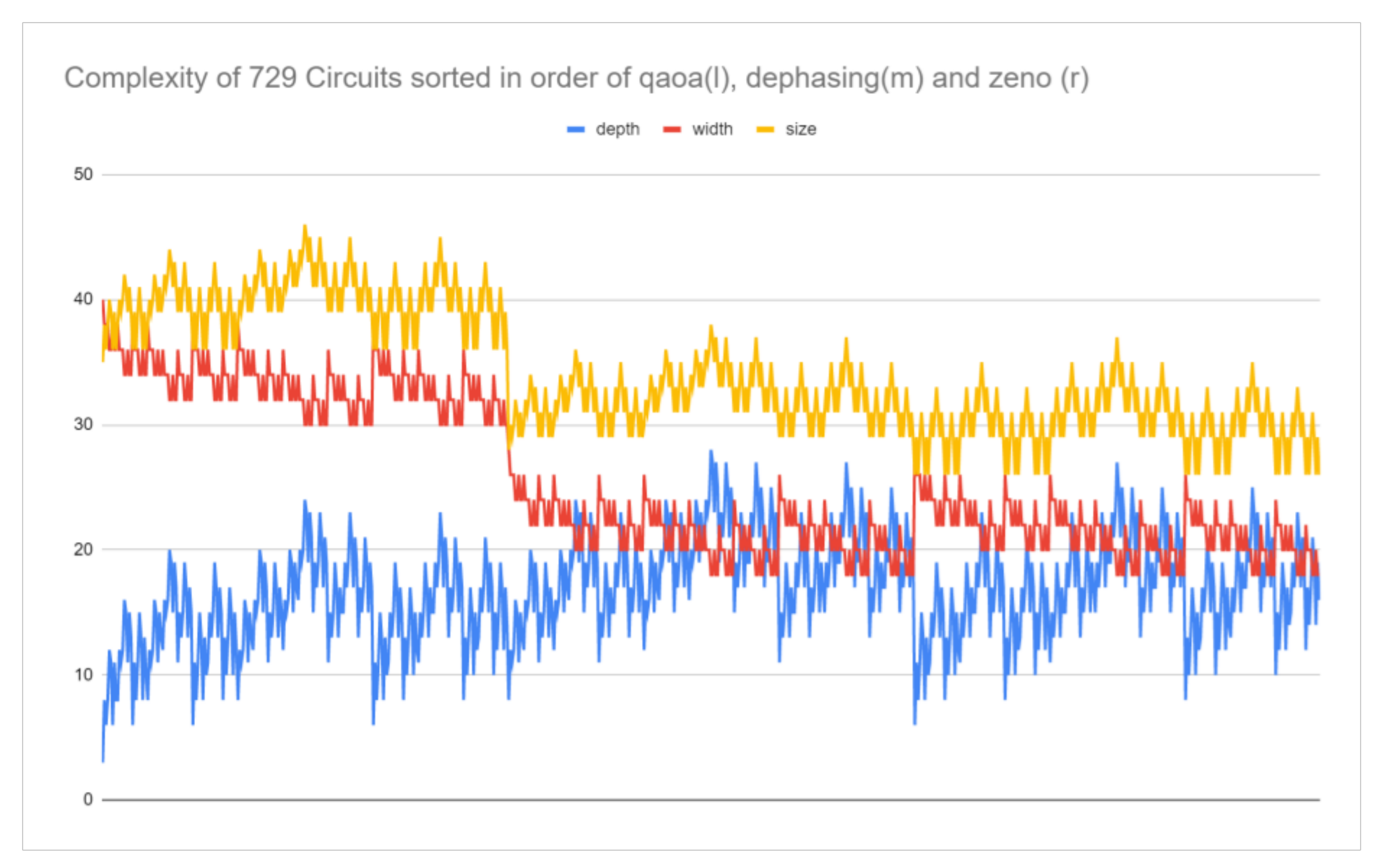}
	\caption{Circuit Structure: Complexity of all circuits}
	\label{fig:complexity729}
\end{figure*}

\begin{figure*}[ht]\centering 
	\includegraphics[width=380pt]{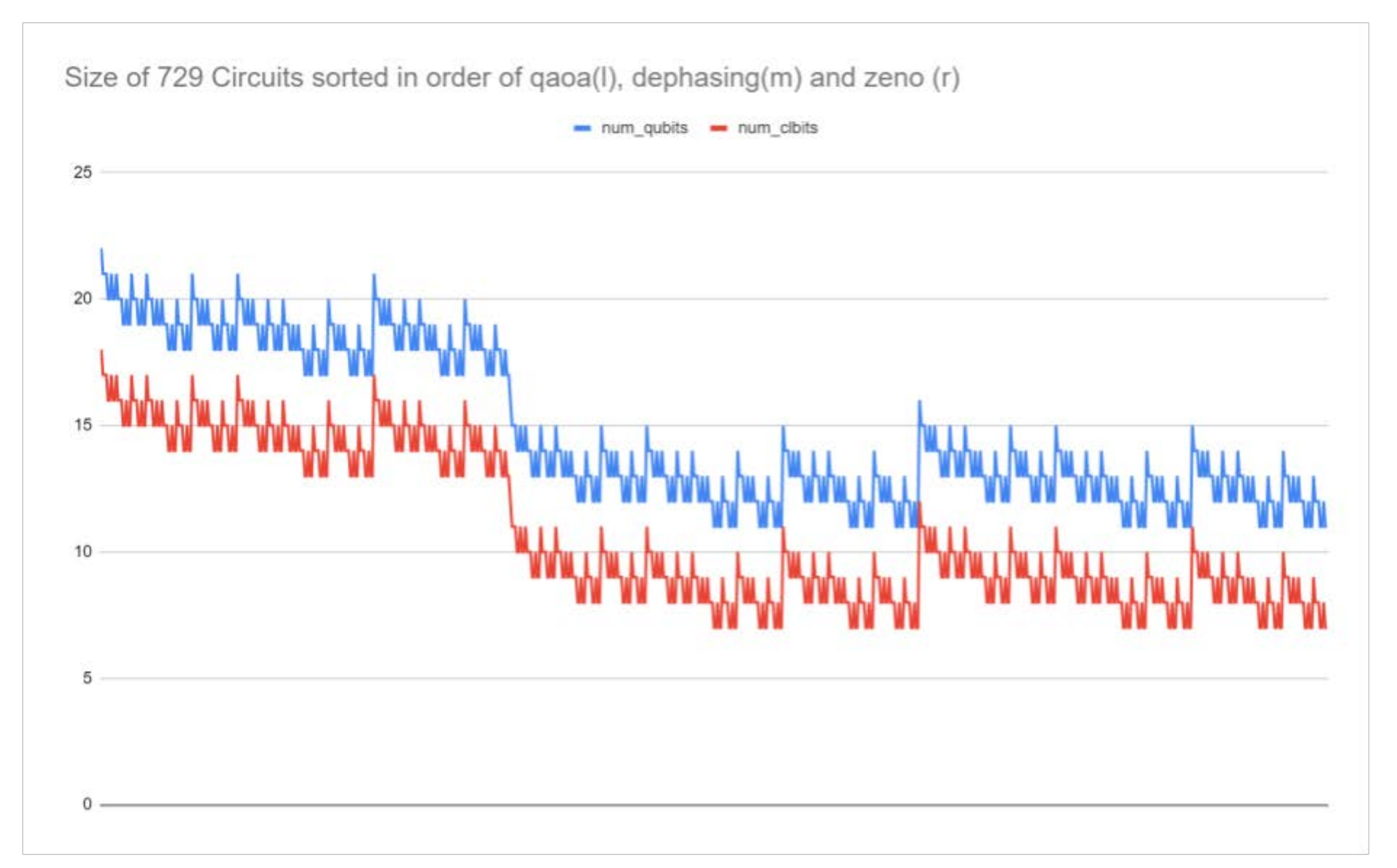}
	\caption{Circuit Structure: Size of all circuits}
	\label{fig:size729}
\end{figure*}

\begin{figure*}[ht]\centering 
	\includegraphics[width=380pt]{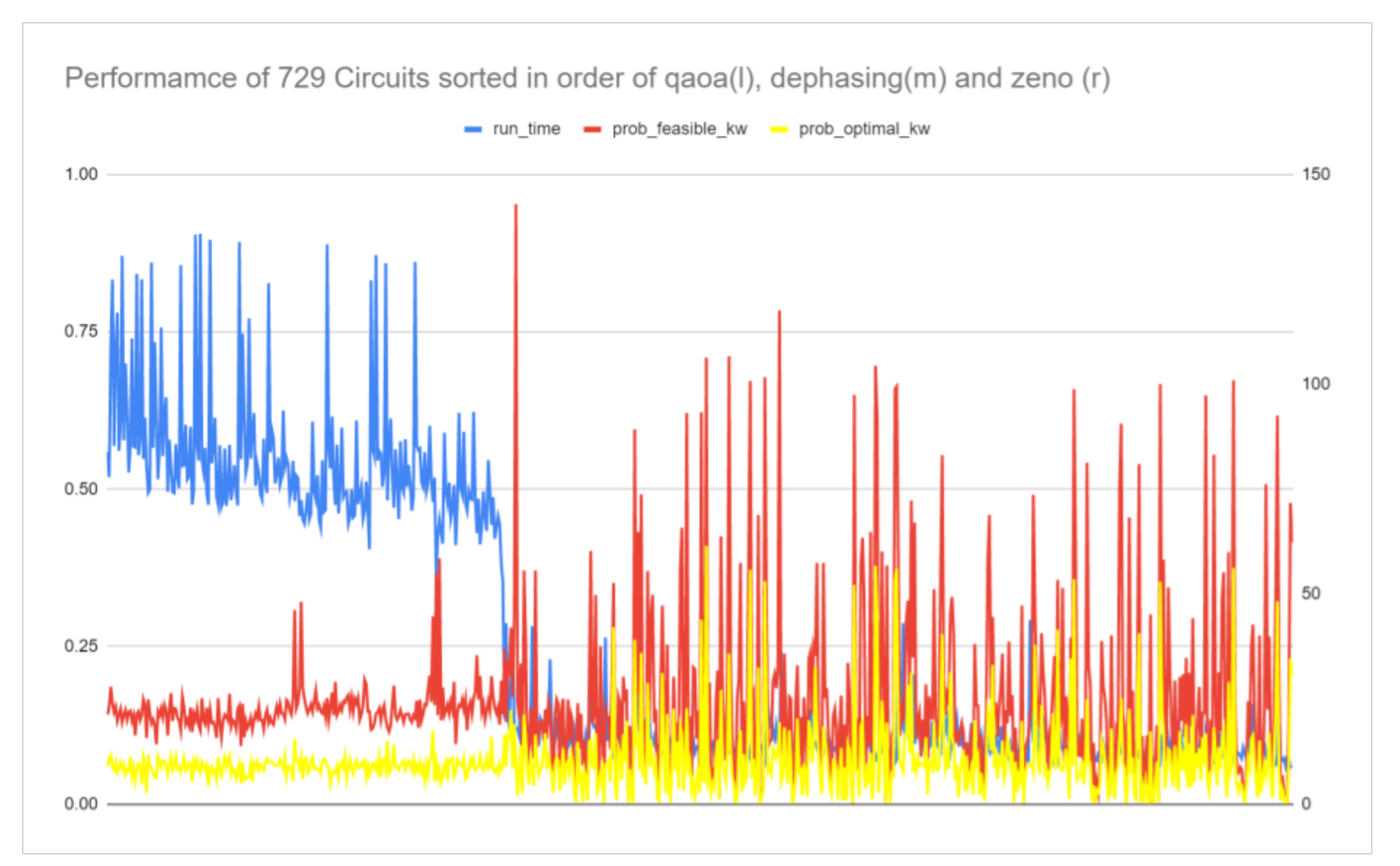}
	\caption{Circuit Structure:  Performance of all circuits}
	\label{fig:performance729}
\end{figure*}

\begin{enumerate}
    \item size and width of circuit using standard qaoa tends to be bigger than circuits using dephasing or zeno only. Here the noticeable level switch is caused by the change around the weight constraint. Please see Figure \ref{fig:complexity729}
    \item In terms of number of qubit and classical bit, the standard qaoa circuits tend to require more because all constraints are squared and added into Lagrange objective, leading to additional slack variables. Please see Figure \ref{fig:size729}
    \item In terms of running time, standard QAOA require longer time on average, partially because it requires a lot more complex gates and super-positions.  Please see Figure \ref{fig:performance729}
    \item In terms of probability of obtaining feasible and optimal solution, the performance of standard qaoa contains less variance while doe not yield high probability. On the other side, dephasing or zeno tends to show high probability in certain specification.  Please see Figure \ref{fig:performance729}
\end{enumerate}

\subsection{Impact of Order of Blocks}

The constraints that are integrated in QAOA are added into the circuits as part of the phase return. After this step, other constraints are appended into the circuit through their own blocks. When adding such blocks into the circuits, one can switch the orders of different constraints leading to slightly different structures.  In general we considered three potential ordering:
\begin{itemize}
\item Natural order: We append the circuits corresponding to the weight constraint first, then those for the positional constraint, followed last by those for the by cargo constraint. This method just order all blocks by a given order without considering the implementation of each constraint
\item Zeno first, Dephasing last: We appended all ZENO blocks after the phase return and then append all dephasing blocks
\item Dephasing first, ZENO last: We appended all dephasing blocks after the phase return and then append all ZENO blocks
\end{itemize}

\begin{figure*}[ht]\centering 
	\includegraphics[width=400pt]{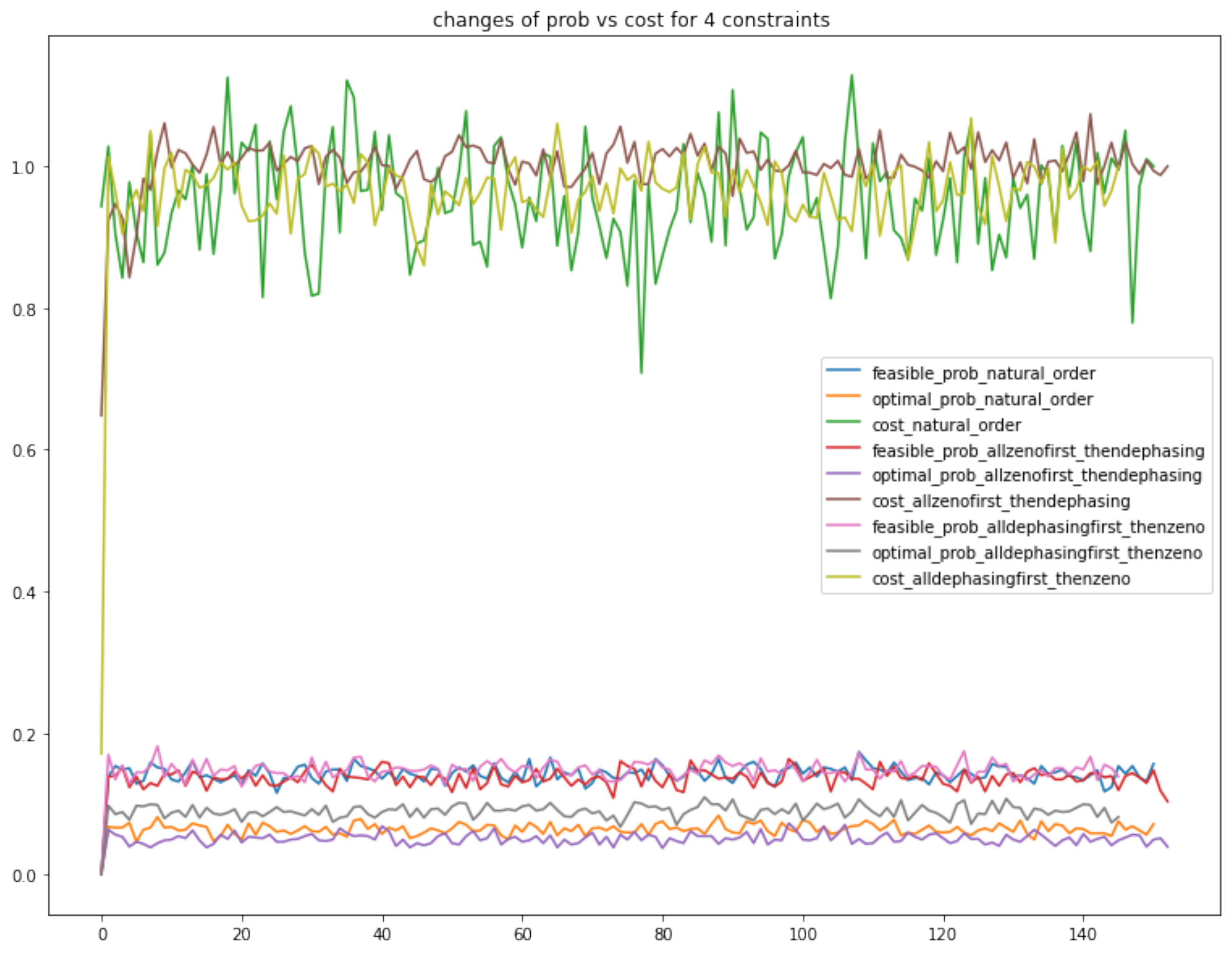}
	\caption{Cost and probability changes over iteration}
	\label{fig:orderingimpact}
\end{figure*}

A quick test is conducted to show the impact, in which 2 constraints are solved in Dephasing and 2 constraints are solved in ZENO. The natural order results in Dephasing-ZENO-Dephasing-ZENO. As shown in Figure \ref{fig:orderingimpact}  it is observed the circuit performance is relatively stable when order changes. 

\subsection{Impact of Initial States}

It is a challenging mission to set initial state for a complex circuit. As we stated in the algorithm it is most theoretically proper to start with a solution which is feasible w.r.t the constraints that are solved by ZENO. To ahieve this goal there are two potential ways:
\begin{itemize}
\item Start with a superposition of all solutions, and then filter away the infeasible solutions w.r.t selected constraints


\item Construct the superposition of solutions w.r.t the selected constraints directly by identifying the convolution matrix. 

\end{itemize}

In practice, we select to pre-run with full states and manually select feasible states for a final run.
\section{Conclusion and Future Work}

An innovative hybrid quantum framework is proposed to solve optimization problems with multiple constraints. This approach involves employing both Ising and non-Ising formulations to represent different parts of the problem. Standard QAOA is applied to the Ising part, while for the non-Ising part, either Quantum Zeno Effect or Penalty Dephasing is employed. This hybrid framework provides a novel method for solving optimization problems through quantum computing. Additionally, we propose a new way to introduce Quantum Zeno Effect into the representation of a selected constraint by frequently measuring the constraint flag. This circuit structure can solve constrained optimization problems with the Zeno effect. Our framework also allows practitioners to specify how to represent each constraint in one of the three alternative structures proposed.

For future work, the algorithm's properties within this framework should be discussed in greater theoretical depth, and a more reliable implementation should be designed for industrial usage. The practical application of the algorithm will rely on advancements in quantum computing, particularly in terms of the number of qubits supported.

\section{Appendix}

\subsection{Probability of retaining initial state for Quantum Zeno effect}

Let's say quantum system state is $|\psi_{\boldsymbol{0} }\rangle $ at time =0, then we apply certain operation under the total Hamiltonian H which results in the wave function evolve to state  
$ |\psi_{\boldsymbol{t} } \rangle $ . The probability that we find the quantum system is still with state $|\psi_{\boldsymbol{0} }\rangle $ even at time t is 
$P_{\boldsymbol{0}}(t) =|\langle \psi_{\boldsymbol{0}}|\psi_{\boldsymbol{t}}\rangle|^2 $. And we also know :
$$|\psi_{\boldsymbol{t} } \rangle = |\psi_{\boldsymbol{0} }\rangle -\frac{i*t}{\hbar} H |\psi_{\boldsymbol{0} }\rangle -\frac{t^2}{2\hbar^2}H^2 |\psi_{\boldsymbol{0} }\rangle + O(t^3) $$
which is a Taylor expansion of $|\psi_{\boldsymbol{t} }\rangle $ and the time derivatives of wave function within it were simplified by Schrödinger equation. Insert this function into $P_{\boldsymbol{0}}(t) =|\langle \psi_{\boldsymbol{0}}|\psi_{\boldsymbol{t}}\rangle|^2 $ we can arrive at:
$$P_{\boldsymbol{0}}(t)= 1-\frac{t^2}{\hbar^2} (\Theta_{\boldsymbol{H}})^2 + O(t^3) $$

$$ {where }\quad
\Theta_{\boldsymbol{H}} = \langle \psi_{\boldsymbol{0}}| H^2|\psi_{\boldsymbol{0}}\rangle - \langle \psi_{\boldsymbol{0}}| H|\psi_{\boldsymbol{0}}\rangle ^2 $$

If we induce the the quantum zeno effect by performing projective measurements N times repetitively to check whether the system is still in its initial state during time t, then each measurement interval is$\quad\varepsilon =t/N$. If after every measurement the system is found in its initial state, the wave function would evolve anew from $|\psi_{\boldsymbol{0} }\rangle $, and the survival probability after the N measurement would be:
$$ P_{\boldsymbol{0}}(t)=(P_{\boldsymbol{0}}(\varepsilon))^N =[1-\frac{\varepsilon^2}{\hbar^2} (\Theta_{\boldsymbol{H}})^2 + O(\varepsilon^3) ^N $$

$$\overrightarrow{large N, small \varepsilon}\quad\ 1-\frac{N*\varepsilon^2}{\hbar^2} (\Theta_{\boldsymbol{H}})^2$$
$$ =1-\frac{t*\varepsilon}{\hbar^2} (\Theta_{\boldsymbol{H}})^2$$

Therefore as we can see when amount the measurement is large enough and measurement interval $ \varepsilon $ is small enough $P_{\boldsymbol{0}}(t)$ is close to 1, which means probability that the system remaining in the initial state is extremely high. 
\subsection{Incomplete measurement and sub-block decomposition of state space for Quantum Zeno effect}
 Consider a quantum system on a finite-dimensional Hilbert space $ \mathbb{H}$, and evolution of this system is governed by the unitary operator U(t)=exp (-iHt), where H is the time-independent Hamiltonian. Also assume measurement is represented by a set of orthogonal projection operators P(which doesn't commute with the Hamiltonian), and the measurement ascertains whether the system is in the subspace       $ P\mathbb{H} $ with less dimensions. If set initial density matrix of the system as $\rho_{\boldsymbol{0}}$, the state at time t is :
$$\rho_{\boldsymbol{t}} = U(t) \rho_{\boldsymbol{0}} U^\dagger(t)$$
If a measurement P is conducted :
$$\rho_{\boldsymbol{t}}\rightarrow P\rho_{\boldsymbol{t}}P=V(t)\rho_{\boldsymbol{0}} V^\dagger(t) \quad {where} \quad V(t) \equiv PU(t)P$$
If conduct N times of P measurements during period of time t:
$$\rho^{(N)}_{\boldsymbol{t}} =V_{\boldsymbol{N}}(t)\rho_{\boldsymbol{0}}V^\dagger_{\boldsymbol{N}}(t) \quad\quad V_{\boldsymbol{N}}(t)\equiv[PU(t/N)P]^N$$
While 
$$ \lim_{N\to\infty}V_{\boldsymbol{N}}(t) =\lim_{N\to\infty}[Pe^{iHt/N}P]^N
= Pe^{-iPHPt}=U_{\boldsymbol{Z}}(t)$$

With above study we can arrive at: Zeno Hamiltionian $H_{\boldsymbol{Z}}\equiv PHP$, and $ U_{\boldsymbol{Z}}(t) $ is actually the unitary operator that governs the system's evolution in subspace $ P\mathbb{H} $

\section*{Acknowledgments}
This work was supported by FinQ Tech Inc. (https://finq.tech/)

\bibliographystyle{unsrt}  
\bibliography{references}

\end{document}